\documentclass[ALICE,manyauthors]{cernphprep}
\usepackage[comma,square,numbers,sort&compress]{natbib}
\usepackage{hyperref}
\usepackage{lineno}
\usepackage{color}
\usepackage{units}
\usepackage{amsmath}
\usepackage{float}
\usepackage[T1]{fontenc}
\usepackage{orcidlink}

\usepackage{xspace}
\usepackage{color}
\usepackage{xcolor}
\usepackage{rotating}
\definecolor{light-gray}{gray}{0.8}

\newcommand{\pt}{\ensuremath{{\it p}_{\rm T}}\xspace}

\newcommand{\cmnt}[1]{}

\newcommand{\GeVc}{\ensuremath{\mathrm{GeV}/c}\xspace}

\newcommand{\dph}{\Delta\varphi }

\newcommand{\pte}{p_{\rm T}^\text{e}}
\newcommand{\ptassoc}{p_{\rm T}^\text{assoc}}

\newcommand{\gevc}{\ensuremath{\mathrm{GeV}/c}\xspace}

\begin{document}%
\begin{titlepage}
\PHyear{2025}
\PHnumber{137}      
\PHdate{14 June}  
%

\title{Medium-induced modification of azimuthal correlations of electrons from heavy-flavor hadron decays with charged particles in Pb--Pb collisions at $\mathbf{\sqrt{s_{\rm{NN}}} = 5.02}$ TeV}
\ShortTitle{Heavy-flavor decay electron--charged particle correlations in Pb--Pb collisions} 

\Collaboration{ALICE Collaboration\thanks{See Appendix~\ref{app:collab} for the list of collaboration members}}
\ShortAuthor{ALICE Collaboration} 

\begin{abstract}
The azimuthal-correlation distributions between electrons from the decays of heavy-flavor hadrons and associated charged particles in Pb--Pb collisions at $\sqrt{s_{\mathrm{NN}}} = 5.02$ TeV are reported for the 0--10\% and 30--50\% centrality classes. This measurement provides access to the jet-like correlation observables in the heavy-flavor sector in Pb--Pb collisions. The analysis is performed for trigger electrons from heavy-flavor hadron decays  with transverse momentum $4 < p_\mathrm{T}^\mathrm{e} < 16~{\rm GeV}/c$, considering associated particles within the transverse-momentum range $1 < p_\mathrm{T}^\mathrm{assoc} < 7$ ${\rm GeV}/c$, and a pseudorapidity difference of $|\Delta\eta|<1$ between the trigger electron and associated particles.
The per-trigger nuclear modification factor ($I_\mathrm{AA}$) is calculated to compare the near- and away-side peak yields to those in pp collisions at $\sqrt{s} = 5.02$ TeV. 
In 0--10\% central collisions, the $I_\mathrm{AA}$ indicates a hint of enhancement of associated-particle yields with $p_\mathrm{T}<3$ GeV/$c$ on the near side, and a suppression of yields with $p_\mathrm{T}>4$ GeV/$c$ on the away side.
The $I_\mathrm{AA}$ for electron triggers from heavy-flavor hadron decays is compared with that for light-flavor and strange-particle triggers to investigate the dependence on different fragmentation processes and parton-medium dynamics, and is found to be the same within uncertainties.

\end{abstract}
\end{titlepage}

\setcounter{page}{2}

%
%
\section{Introduction}
\label{sec:Intro}
In hadronic collisions, heavy quarks are produced in hard-scattering processes, with large squared-momentum transfer, $Q^2$~\cite{Kniehl:2007erq,Cacciari:2003uh,Kniehl:2005mk,Cacciari:2003zu}. High-energy partons produced in such hard scatterings radiate gluons, generating a parton shower. The shower hadronizes, producing a collimated spray of particles called a jet. Jet-reconstruction algorithms~\cite{Cacciari:2008gp,Cacciari:2011ma} can be used to study their production and substructure, which provide stringent tests of perturbative quantum chromodynamics (pQCD) calculations. 
Two-particle angular correlations are a complementary method to jet-reconstruction measurements for characterizing jet properties, especially at low $\pt$~\cite{ALICE:2011gpa,ALICE:2016gso}. The typical features of a two-particle angular-correlation distribution of trigger particles with associated charged particles are a ``near-side'' (NS) peak at $(\Delta\varphi, \Delta\eta) \approx (0,0)$ and an ``away-side'' (AS) peak at $\Delta\varphi \approx \pi$, which extends over a wide pseudorapidity range~\cite{ALICE:2016clc,ALICE:2019oyn,ALICE:2023kjg}. 

At leading order (LO), heavy quarks are produced through the processes $gg, q\bar{q} \rightarrow c\bar{c}$ and $gg, q\bar{q} \rightarrow b\bar{b}$, which lead to a back-to-back configuration in azimuth. In the two-particle angular-correlation distribution, this results in a NS peak from particles emerging from the fragmentation of the same parton that produced the trigger particle~\cite{Mangano:1991jk}. The AS peak is associated with the fragmentation of the other outgoing parton produced in the hard scattering~\cite{Mangano:1991jk}, which at LO is the partner quark in the heavy-quark pair.
Next-to-leading order (NLO) processes, such as gluon radiation, gluon splitting, or flavor excitation~\cite{Norrbin:2000zc} can alter the LO topology of the correlation shapes~\cite{Mangano:1991jk, Norrbin:2000zc, ALICE:2023kjg}. Angular correlation measurements with a heavy-flavor trigger particle facilitate the study of heavy-flavor jet properties. By varying the $\pt$ intervals of the trigger and associated particles, the details of jet fragmentation can be investigated, such as the jet angular profile and the momentum distribution of the particles produced in the hard-parton fragmentation. 

In ultrarelativistic heavy-ion collisions, a deconfined phase of strongly-interacting matter is produced, known as quark--gluon plasma (QGP)~\cite{Ipek:2018lhm, Bazavov:2011nk, Borsanyi:2010cj, Bazavov:2018mes, Bazavov:2009zn}. Experiments at the CERN Large Hadron Collider (LHC)~\cite{Muller:2012zq, ALICE:2022wpn,CMS:2024krd} and the Brookhaven Relativistic Heavy Ion Collider (RHIC)~\cite{Arsene:2004fa, Back:2004je, Adams:2005dq, Adcox:2004mh} have extensively studied the properties of the QGP, characterizing it as a near-perfect liquid with very low specific shear viscosity-to-entropy ratio. One of the signatures of QGP is the parton energy loss, where high-energy partons traversing this medium lose energy through medium-induced gluon radiation and collisions with medium constituents~\cite{Baier:2000mf, Dokshitzer:2001zm, Armesto:2003jh, Wicks:2007am, Zhang:2003wk, Adil:2006ra}. This phenomenon, 
referred to as ``jet quenching'', depends on the path length traversed by the parton. In this process, the particle distribution within the jet undergoes modification, changing the internal structure of the jet, thus affecting the fragmentation function~\cite{Qin:2015srf,Blaizot:2014ula}. 
A previous measurement of the jet fragmentation function shows modifications in Pb--Pb collisions compared to pp collisions, with an excess of low-$\pt$ particles in inclusive jets~\cite{ATLAS:2018bvp,CMS:2014jjt,Cao:2022odi,Li:2024pfi}.
High-momentum partons propagating through the QGP medium can also lead to modifications in the QGP itself due to the injection of energy and momentum lost by the jet into the plasma~\cite{Qin:2009uh}. This yields a correlation between the bulk dynamics of the medium and the jet~\cite{Casalderrey-Solana:2020rsj,Yang:2025dqu,Yang:2022nei, Chen:2023dzi}. Measurements of heavy-flavor jets and particle distributions within jets can serve to constrain parton energy-loss mechanisms and probe how the lost energy is redistributed among other partons and subsequent particles emerging from the collision. Theoretical models predict that energy loss depends on the quark mass and the color charge of the parton~\cite{Wicks:2007am, Armesto:2003jh,Djordjevic:2004nq,Dokshitzer:2001zm}. Heavy quarks are expected to lose less energy than light quarks~\cite{Wicks:2007am, Armesto:2003jh,Djordjevic:2004nq}, which is corroborated by measurements of the nuclear modification factor ($R_{\rm{AA}}$) of charm and beauty hadrons~\cite{ALICE:2018yau,ALICE:2018lyv,ALICE:2012ab,CMS:2017qjw,ALICE:2022tji,ALICE:2022iba, ATLAS:2021xtw}, and jets containing heavy-flavor particles~\cite{CMS:2013qak, CMS:2019jis, ATLAS:2022agz, CMS:2022btc}. 

Measurements of two-particle angular correlations using light-flavor hadrons as trigger particles are employed by the ALICE Collaboration~\cite{ALICE:2011gpa,ALICE:2016gso,ALICE:2022rau} to study jet quenching in Pb--Pb collisions. These measurements indicate a strong suppression of the per-trigger away-side peak yield at high associated-particle $\pt$ in central Pb--Pb collisions compared to pp collisions. The absence of suppression on the NS, combined with the observed suppression on the AS, is consistent with the away-side parton experiencing a larger average path length in the medium. An enhancement of the low-$\pt$ associated-particle yield on the near side is also observed.
The ALICE Collaboration has also measured charged-particle jets recoiling from a high-$\pt$ parton~\cite{ALICE:2023jye,ALICE:2023qve}, and observed a strong medium-induced enhancement of recoil jets at low $p_{\rm{T,jet}}$ in Pb--Pb collisions compared to pp collisions. This enhancement potentially arises from the response of the QGP medium to jets, as suggested by calculations that include such a medium response~\cite{Casalderrey-Solana:2014bpa}. 

Measurements of the angular correlations using heavy-flavor hadrons as trigger particles, and studying their peak yield modifications, can expand the investigation of jet transport and QGP dynamics in the heavy-flavor sector, allowing for further exploration of the parton mass dependence of parton–medium interactions.
The PHENIX and STAR Collaborations at RHIC conducted studies of angular correlations of electrons from heavy-flavor hadron decays with charged hadrons~\cite{PHENIX:2010cfl} and angular correlations of D mesons with charged hadrons~\cite{STAR:2019qbf} in Au--Au collisions at $\sqrt{s_{\rm{NN}}}=200$ GeV. Although these correlation measurements are statistically limited, they indicate modifications of the NS and AS correlation peaks in central Au--Au collisions compared to pp collisions, suggesting QGP effects. The CMS Collaboration investigated the modification of the jet profile initiated by beauty quarks (b jets)~\cite{CMS:2022btc} and charm-quark diffusion with respect to the jet axis~\cite{CMS:2019jis}. The measurement illustrates a redistribution of the $\pt$ of jet constituents from small to large distances from the jet axis. A large-distance enhancement is observed in Pb--Pb collisions, more pronounced for b jets than for inclusive jets, indicating mass-dependent interactions in the QGP.

This article presents a jet-like correlation measurement in Pb--Pb collisions at LHC energies using a particle from the heavy-flavor sector as the trigger particle. The study is based on azimuthal-correlation distributions between electrons from heavy-flavor hadron decays and associated charged particles in Pb--Pb collisions at $\sqrt{s_{\rm NN}} = 5.02$ TeV, conducted using the ALICE detector. Since the electrons originate from decays of charm and beauty hadrons, which in turn result from the fragmentation of charm and beauty quarks, the correlation measurement provides access to the study of heavy-quark fragmentation.
The correlation distributions are studied for various $p_{\rm T}$ intervals of both trigger electrons and associated charged particles to explore the $\pt$ dependence of the correlation shapes.
The results are compared with those measured in pp collisions at the same center-of-mass energy~\cite{ALICE:2023kjg} to investigate the effects of the QGP medium on the particle distribution within the heavy-flavor jet. 

 The article is organized as follows. The ALICE apparatus, its main detectors used in the analyses, and the data samples are reported in Sec.~\ref{chp:detector}. The complete analysis procedure is described in Sec.~\ref{chp:analysis}. The systematic uncertainties associated with the measurements are discussed in Sec.~\ref{chp:systematics}. The analysis results are presented and discussed in Sec.~\ref{chp:results}. The article is briefly summarized in Sec.~\ref{chp:summary}.

\section{Experimental apparatus and data sample}
\label{chp:detector}
The measurement is performed with the ALICE detector at the LHC. A complete description of the ALICE detector is detailed in Refs.~\cite{ALICE:2014sbx,Aamodt:2008zz}. The central barrel of the detector is located at midrapidity ($|\eta| < 0.9$) and is embedded within a cylindrical solenoid that produces a magnetic field of 0.5 T parallel to the direction of the beam. The main detectors used for this analysis are the Inner Tracking System (ITS), the Time Projection Chamber (TPC), the electromagnetic calorimeters (EMCal and DCal), the V0, and the Zero Degree Calorimeters (ZDC). 
The ITS~\cite{ALICE:2010tia} is used for the reconstruction of the primary vertex and charged-particle tracking. It consists of six cylindrical layers of silicon detectors, with the innermost two layers composed of Silicon Pixel Detectors (SPD), the middle two made of Silicon Drift Detectors, and the outer layers composed of Silicon Strip Detectors. The TPC~\cite{Alme:2010ke} is the core of the ALICE tracking system, providing three-dimensional information of particle momenta as well as specific energy loss (d$E$/d$x$) measurements for the purpose of charged-particle identification~\cite{Bethe:1930ku,ALEPH:1994ayc,ALEPH:1994dex}. The EMCal ($|\eta| < 0.7$ and $80^{\circ} < \varphi < 187^{\circ}$) and DCal ($0.22 < |\eta| < 0.7$ and $260^{\circ} < \varphi < 320^{\circ}$, $|\eta| < 0.7$ and $320^{\circ} < \varphi < 327^{\circ}$) detectors~\cite{Cortese:1121574,Allen:2010stl} are used for electron identification and event triggering. 
The EMCal and DCal detectors will be collectively referred to as EMCal for the remainder of this paper. The V0 system is used for centrality estimation~\cite{ALICE:2013axi} and is composed of two scintillator arrays, V0A and V0C, which cover the pseudorapidity ranges of $2.8 < \eta < 5.1$ and $-3.7 < \eta < -1.7$, respectively. Simultaneous signals in both the V0A and V0C detectors define the minimum-bias interaction (MB) trigger. The ZDC~\cite{Cortese:2019nnv}, located at 112.5 m on both sides of the interaction point, are utilized for rejecting electromagnetic interactions and beam-induced background in Pb--Pb collisions.

The results presented in this paper are obtained from Pb--Pb collisions at the center-of-mass energy of $\sqrt{s_{\rm{NN}}}=5.02$ TeV recorded during the LHC Run 2 and selected using the MB trigger of the 2018 run exploiting the V0 detector. Two centrality percentile classes~\cite{ALICE:2018tvk} are used: 0--10\%, classified as the most-central collision events, and 30--50\%, categorized as semicentral collision events. Pileup events with two or more primary vertices are rejected via an algorithm that detects multiple vertices reconstructed by the SPD. For the purpose of maintaining uniform acceptance, only collisions with a reconstructed primary vertex within $\pm$ 10 cm from the nominal interaction point ($z=0$) along the beam axis are selected. The number of events analyzed is $41.5\times10^{6}$ ($62.3\times10^{6}$) in 0--10\% (30--50\%) central (semicentral) Pb--Pb collisions, corresponding to an integrated luminosity of approximately $54~\mu\mathrm{b}^{-1}$ ($41~\mu\mathrm{b}^{-1}$)~\cite{ALICE:2018tvk}, respectively.
\section{Analysis procedure}
\label{chp:analysis}

The analysis is performed by evaluating the two-particle azimuthal-correlation distributions between trigger electrons (and positrons) from heavy-flavor hadron decays~\cite{ParticleDataGroup:2024cfk} and associated charged primary particles. Particles are defined as primary if they have a mean proper lifetime $\tau > 1$ cm/$c$ and are produced directly in the collision. A particle is also classified as primary if its parent has $\tau < 1$ cm/$c$ and the parent originates from a decay chain arising from the collision. Secondary particles are those that do not meet the criteria for being classified as primaries~\cite{ALICE-PUBLIC-2017-005}.
Effects on the measured correlation distribution, induced by the detector’s limited acceptance and local inhomogeneities, are corrected using an event-mixing technique. The correlation distributions are normalized by the number of electron triggers and corrected for the associated-particle tracking efficiency. The contamination from hadrons misidentified as electrons, electrons from background sources, as well as secondary associated particles, is subtracted statistically from the measured distributions, as discussed below.  
The per-trigger peak yields are obtained after subtracting the contribution from uncorrelated pairs, referred to as baseline. The following section details the analysis procedure, which largely follows the procedure used in the measurement of angular correlations of electrons from heavy-flavor hadron decays and charged hadrons in pp and p--Pb collisions at $\sqrt{s_\mathrm{NN}}=5.02$ TeV~\cite{ALICE:2023kjg}.  Both positrons and electrons from heavy-flavor hadron decays will be collectively referred to as ``electrons" for the remainder of this paper. 

\subsection{Electron identification and associated-particle reconstruction}
Electron tracks 
with transverse momentum within $4 < \pte < 16$ GeV/$c$ and within a pseudorapidity interval of $|\eta| < 0.6$ are considered. Track quality selections are applied by requiring the same selection criteria as defined in Ref.~\cite{ALICE:2019nuy, ALICE:2023kjg}, and at least one hit in the SPD.
To reject secondary electrons, which are produced from interactions with the detector material or from the weak decays of long-lived particles, a distance of closest approach requirement is imposed using the same selection as in Ref.~\cite{ALICE:2023kjg}. Particle identification is implemented by imposing a d$E$/d$x$ selection on the TPC track and via the energy deposition in the EMCal detector. The discriminant variable for the TPC electron identification is the number of standard deviations ($n\sigma_\text{e}^\text{TPC}$) of the measured d$E$/d$x$ from the expected value for electrons obtained from the Bethe-Bloch parameterization~\cite{Bethe:1930ku}. An asymmetric selection of $-1 < n\sigma_\text{e}^\text{TPC} < 3$ is applied due to the larger hadron background contamination in the region of negatively valued $n\sigma_\text{e}^\text{TPC}$. The ratio of the energy ($E$) deposited in the EMCal to the momentum ($p$) measured by the TPC provides another method of electron identification. Electrons produce an electromagnetic shower in the EMCal and deposit their total energy $E$.  Because of their small mass relative to their momentum, their $E/p$ ratio is very close to 1. The electron identification with the EMCal is thus performed by selecting candidates with $0.8 < E/p < 1.2$. In addition, the shape of the electromagnetic shower in the EMCal is used for electron identification and to reduce hadron contamination. The shower shape is characterized by the eigenvalues of the dispersion matrix of the shower shape ellipse defined by the energy distribution within the EMCal cluster~\cite{ALICE:2022qhn,AWES1992130}. A long-axis selection of $0.02 < \sigma^{2}_\mathrm{long} < 0.9$ is used in the analysis.

Charged primary particles~\cite{ALICE-PUBLIC-2017-005} (defined as "associated particles") are correlated with the trigger particle within $|\eta| < 0.8$. The reconstruction of charged primary particles follows the same criteria as those in Ref.~\cite{ALICE:2023kjg}. Track quality selections include requirements on the minimum number of crossed TPC pad rows and the minimum ratio of crossed TPC pad rows to findable clusters. Additionally, a distance of closest approach (DCA) requirement of $\mathrm{DCA_{xy}}<0.5$ cm in the transverse plane and $\mathrm{DCA_{z}}<1.0$ cm in the longitudinal direction is imposed to reject a large fraction of secondary particles. 

The associated particles are required to have a smaller \pt than the trigger electron when selecting trigger–associated pair candidates. This follows the same requirement used in the pp and p--Pb analyses~\cite{ALICE:2023kjg}, and introduces a kinematic bias in the \pt intervals where $p_\mathrm{T}^\mathrm{trig}$ and $p_\mathrm{T}^\mathrm{assoc}$ overlap. This bias is reproduced by simulations and model predictions.

\subsection{Azimuthal-correlation distribution and mixed-event correction}

A two-dimensional correlation distribution, $C(\Delta\eta,\Delta\varphi)$, is evaluated as a function of the azimuthal-angle difference ($\Delta\varphi$) and the pseudorapidity difference ($\Delta\eta$) between trigger and associated particles. This distribution is computed for trigger $p_\text{T}$ intervals $4 < p_\text{T}^\text{e} < 12$ GeV/$c$, as well as $4 < p_\text{T}^\text{e} < 7$ GeV/$c$ and $7 < p_\text{T}^\text{e} < 16$ GeV/$c$, in each case for five associated $p_\text{T}$ intervals between 1 and 7 GeV/$c$: $1 < p_\text{T}^\text{assoc} < 2$ GeV/$c$, $2 < p_\text{T}^\text{assoc} < 3$ GeV/$c$, $3 < p_\text{T}^\text{assoc} < 4$ GeV/$c$, $4 < p_\text{T}^\text{assoc} < 5$ GeV/$c$, and $5 < p_\text{T}^\text{assoc} < 7$ GeV/$c$. These trigger and associated \pt intervals are selected to facilitate direct comparisons to the equivalent measurement in pp collisions~\cite{ALICE:2023kjg}. In addition, the electron $p_\text{T}$ is limited to $16~\mathrm{GeV}/c$ to reduce the contribution from $W$- and $Z$-boson decays~\cite{ALICE:2023xiu,ALICE:2016rzo}.
The correlation distributions are corrected for pair acceptance and detector inhomogeneities for each $p_\text{T}$ interval using a mixed-event technique as described in Ref.~\cite{ALICE:2013snk}. Equation~\ref{eqn:Mixed_Event} shows the mixed-event correction procedure, where $S(\Delta\eta,\Delta\varphi)$ represents the two-dimensional correlation distribution after correction by the normalized mixed-event distribution $\displaystyle{\frac{1}{\beta}}ME(\Delta\eta,\Delta\varphi)$. 

\begin{equation}
    \frac{\text{d}^{2}N}{\text{d}\Delta\eta\text{d}\Delta\varphi} \equiv S(\Delta\eta,\Delta\varphi) = \beta \times \frac{C(\Delta\eta,\Delta\varphi)}{ME(\Delta\eta,\Delta\varphi)}
\label{eqn:Mixed_Event}
\end{equation}

The mixed-event correlation distribution, $ME(\Delta\eta,\Delta\varphi)$, is constructed by correlating trigger electrons with associated charged particles from different collisions that have similar multiplicity and primary-vertex position along the beam axis as the events containing the electron. The resulting mixed-event correlation distribution is characterized by a triangular shape along the $\Delta\eta$ axis, due to the limited $\eta$ coverage, and is approximately flat in $\Delta\varphi$. In the region $(\Delta\eta,\Delta\varphi) \approx (0,0)$, the trigger and associated particles cross the same detector elements, therefore this region of the correlation distribution is not significantly affected by detector inhomogeneities. This property is used to obtain a normalization factor, $\beta$, for $ME(\Delta\eta,\Delta\varphi)$. The $\beta$ factor is defined as the average number of correlated pairs within $-0.2 < \Delta\varphi < 0.2$ and $-0.07 < \Delta\eta < 0.07$ in the mixed-event correlation. To reduce the effect of statistical fluctuations, particularly at large $|\Delta\eta|$, the mixed-event corrected two-dimensional distribution, $S(\Delta\eta,\Delta\varphi)$, is integrated over the pseudorapidity range of $|\Delta\eta| < 1$ to obtain a one-dimensional distribution $S(\Delta\varphi)$.

\subsection{Background subtraction}

To obtain the final electron correlation distribution, two types of background must be removed from the electron signal: hadrons incorrectly identified as electrons, and electrons not from heavy-flavor hadron decays. To estimate the contribution of the hadron contamination in the trigger-electron sample, a sample consisting of hadrons is obtained by applying a selection of $n\sigma_\text{e}^\text{TPC} < -3.5$ on the reconstructed tracks. The corresponding $E/p$ distribution from these hadrons is normalized by the electron-candidate $E/p$ in the region $0.25 < E/p < 0.5$, which is displaced sufficiently from the electron peak, in order to match the hadron contamination in the electron-candidate $E/p$ distribution. The normalized hadron $E/p$ is subtracted from the electron-candidate $E/p$ distribution to obtain the hadron-contamination corrected $E/p$ distribution. The number of hadrons from the normalized hadron distribution in the region of $0.8 < E/p < 1.2$ reflects the contamination ($N_\mathrm{had.contam.}$) in the electron-candidate sample. This procedure follows a method similar to that described in Refs.~\cite{ALICE:2019bfx, ALICE:2023kjg}. The average contamination is 5\% (2.5\%) at $\pt^{\rm{e}} = 4$ GeV/$c$ in the 0--10\% (30--50\%) centrality interval, and 22\% (18\%) at $p_\mathrm{T}^\mathrm{e} = 16$ GeV/$c$ in the 0--10\% (30--50\%) centrality interval. To remove the hadron contamination in the $\Delta\varphi$ correlation distribution, an additional distribution is obtained by selecting trigger particles with $n\sigma_\text{e}^\text{TPC}< -3.5$. This hadron-triggered correlation distribution is normalized by $N_\mathrm{had.contam.}$ and subtracted from the electron-candidate correlation distribution to obtain the hadron-contamination corrected $\Delta\varphi$ distribution. The electron sample after hadron contamination removal will be referred to as inclusive electrons (Incl.e).

The resulting azimuthal-correlation distributions also contain a background contribution from electrons not originating from heavy-flavor hadrons. The signal electron sample for this analysis originates from heavy-flavor hadron semileptonic decays, which produce a single electron. The primary sources of the background electrons in the $p_\text{T}^\text{e}$ regions studied are electron-positron ($\mathrm{e}^{+}\mathrm{e}^{-}$) pairs from Dalitz decays of neutral mesons ($\pi^{0} \rightarrow \gamma~\mathrm{e}^{+}\mathrm{e}^{-}$ and $\eta \rightarrow \gamma~\mathrm{e}^{+}\mathrm{e}^{-}$) and photon conversions ($\gamma \rightarrow \mathrm{e}^{+}\mathrm{e}^{-}$) within the detector material. Other potential background sources of electrons such as from kaon, light vector meson, and quarkonium decays or from Drell-Yan processes are negligible in the $p_\text{T}^\text{e}$ intervals selected for this analysis~\cite{ALICE:2012mzy}. To identify these background electrons, unlike-sign (ULS) pairs are built and their invariant mass ($m_\mathrm{e^{+}e^{-}}$) is reconstructed using a technique described in Refs.~\cite{ALICE:2019nuy, ALICE:2016mpw}, resulting in a signal peak at $m_\mathrm{e^{+}e^{-}}=0$ GeV/$c^{2}$. Partner electrons are selected using similar but looser track-quality and electron-identification criteria than those applied to the trigger tracks, in order to optimize the efficiency of reconstructing the partner (Refs.~\cite{ALICE:2016mpw, ALICE:2018yau}). Trigger electrons are paired with a partner electron identified with the TPC selection $-3 < n\sigma_\text{e}^\text{TPC} < 3$ to maximize the selection efficiency. The invariant-mass distribution of combinatorial electron pairs is estimated by reconstructing like-sign electron (LS) pairs. Unlike in the equivalent analysis in pp and p--Pb collisions~\cite{ALICE:2023kjg}, the LS invariant-mass distribution in Pb--Pb collisions does not perfectly match the ULS distribution. This discrepancy arises due to a combinatorial effect, where for a given number of positive and negative tracks, the number of ULS pairs ($\rm{e}^{\pm}\rm{e}^{\mp}$) exceeds the number of LS pairs ($\rm{e}^{+}\rm{e}^{+}$ and $ \rm{e}^{-}\rm{e}^{-}$).
For $4 < p_\mathrm{T}^\mathrm{e} < 12$ GeV/$c$, the LS distribution is thus scaled by a normalization factor of 1.045 to match the ULS distribution in the invariant-mass region $0.2 < m_\mathrm{e^{+}e^{-}} < 0.25$ GeV/$c^{2}$, a region which is sufficiently displaced from the signal peak. The corresponding scale factors for trigger ranges $4 < p_\mathrm{T}^\mathrm{e} < 7$ and $7 < p_\mathrm{T}^\mathrm{e} < 16$ \gevc~are 1.038 and 1.114, respectively. Electrons matched with ULS and LS partners are used to construct azimuthal-correlation distributions, $S(\Delta\varphi)^\text{ULS}$ and $S(\Delta\varphi)^\text{LS}$, respectively. 
 Background contributions are removed by subtracting the LS distribution from the ULS distribution, considering only electron pairs with $m_\mathrm{e^{+}e^{-}} < 0.14$ GeV/$c^{2}$, which captures the majority of the signal from photon conversions and Dalitz decays. The resulting distribution, $(S(\Delta\varphi)^\text{ULS} - S(\Delta\varphi)^\text{LS})$, is then corrected for the partner-finding efficiency, referred to as the tagging efficiency $(\epsilon_\text{tag})$, which is estimated from Monte Carlo simulations following a similar procedure as in Refs.~\cite{ALICE:2016mpw,ALICE:2015zhm}. 
The resulting correlation distribution between electrons from heavy-flavor hadron decays and charged particles, ($S(\Delta\varphi)^\mathrm{c,b \rightarrow e}$), is calculated as

\begin{equation}
    S(\Delta\varphi)^\mathrm{c,b \rightarrow e} = S(\Delta\varphi)^\text{Incl.e} - \frac{1}{\epsilon_\text{tag}}\bigl[S(\Delta\varphi)^\text{ULS}-S(\Delta\varphi)^\text{LS}\bigr].
\label{eqn:Photonic_Correction}
\end{equation}

The Monte Carlo sample used for determining the tagging efficiency is obtained using HIJING v1.36~\cite{Wang:1991hta}, from now on referred to as HIJING. The particle propagation through the ALICE apparatus is simulated using GEANT3~\cite{Brun:1073159}. To improve the statistical precision of the tagging efficiency, additional $\pi^{0}$ and $\eta$ mesons generated with PYTHIA 6.4.25~\cite{Sjostrand:2006za} are embedded into the simulated sample, extending their spectra to a high-$p_\text{T}$ range. For collisions in the 0--10\% (30--50\%) centrality interval, the tagging efficiency is $57\pm1$\% ($61\pm1$\%) at $p_\mathrm{T}^\mathrm{e} = 4$ GeV/$c$, and approximately $65\pm1$\% ($68\pm1$\%) at $p_\mathrm{T}^\mathrm{e} = 7$ GeV/$c$. For $p_\mathrm{T}^\mathrm{e} > 8$ GeV/$c$, the tagging efficiency is approximately constant with a value of $70\pm1$\% for both centrality classes. Other sources of electrons, such as from $\mathrm{J}/\psi$, kaon, and light vector meson ($\rho, \omega$, and $\varphi$) decays, are negligible in the $p_\text{T}^\text{e}$ ranges used in this analysis~\cite{ALICE:2012mzy}.

The $S(\Delta\varphi)^\mathrm{c,b \rightarrow e}$ distribution is corrected for detector inefficiencies in reconstructing associated particles and for contamination from secondary particles within the associated-particle sample. The associated-particle reconstruction efficiency for charged primary particles is obtained from simulated Pb--Pb collisions produced with the HIJING~\cite{Wang:1991hta} event generator. In the interval $1 < p_\text{T}^\text{assoc.}< 7$ GeV/$c$, for 0--10\% (30--50\%) collisions, the efficiency is 82--85\% (86--87\%). This simulated sample is also used to estimate the contamination of secondary particles in the associated-particle sample. The secondary contamination is 2--4\% (1--4\%) for 0--10\% (30--50\%) collisions. The efficiency for trigger-electron reconstruction and selection is approximately constant within the measurement interval, so no correction is needed. 
The resulting azimuthal-correlation distribution is normalized by the number of trigger electrons from heavy-flavor hadron decays ($N_\mathrm{c,b \rightarrow e}$), which is obtained from the number of inclusive electrons ($N_\text{Incl.e}$) and the number of electrons ULS and LS pairs using

\begin{equation}
    N_\mathrm{c,b \rightarrow e} = N_\text{Incl.e} - \frac{1}{\epsilon_\text{tag}}(N_\text{ULS}-N_\text{LS}).
\label{eqn:N_HFe}
\end{equation}

\subsection{Baseline estimation and fit function for the azimuthal-correlation distributions}
In the high-multiplicity environment of Pb--Pb collisions, the per-trigger normalized azimuthal-correlation distribution is affected by a significant background contribution from particle pairs not originating from the same hard scattering process. While this background is typically flat in $\Delta \varphi$ and lies beneath the physically correlated peaks, the presence of collective motion in the medium introduces an additional modulation due to the anisotropic flow of electrons from heavy-flavor hadron decays and the flow of associated charged particles. In this situation, this contribution can be estimated by a baseline function, which is discussed in Ref.~\cite{ALICE:2011svq} and defined as

\begin{equation}
    B(\Delta \varphi) = b(1+2\sum\limits_{n} v_{n}^\mathrm{c,b \rightarrow e} v_{n}^\text{assoc}\cos{(n\Delta \varphi)}),
\label{eqn:Baseline_function}
\end{equation}

where $b$ is a free parameter which determines the height of the baseline function and reflects the magnitude of the correlation background. The terms $v_{n}^\mathrm{c,b \rightarrow e}$ and $v_{n}^\text{assoc}$ are the n-th order harmonic flow coefficients of electrons from heavy-flavor hadron decays and charged associated particles, respectively. For this study, only the $v_{2}$ and $v_{3}$ flow coefficients are considered and the higher-order flow coefficients (for $n \geq 4$), accounting for a negligible contribution to $B(\Delta \varphi)$, are neglected. Measurements by the ATLAS experiment of the second and third flow coefficients for muons produced from heavy-flavor hadron decays at midrapidity in Pb--Pb collisions at $\sqrt{s_{\rm{NN}}} = 5.02$ TeV in the 0--10\% and 30--40\% centrality classes~\cite{ATLAS:2020yxw} are used for the $v_{2}^\mathrm{c,b \rightarrow e}$ and $v_{3}^\mathrm{c,b \rightarrow e}$ parameters. The ATLAS $v_{n}^{\mathrm{c,b \rightarrow \mu}}$ is used in place of $v_{n}^\mathrm{c,b \rightarrow e}$ values because measurements of inclusive electrons from heavy-flavor hadron decays are not available in the $p_\mathrm{T}$ and centrality intervals considered in this analysis. Flow coefficients of charged particles $v_{2}^\text{ch. part.}$ and $v_{3}^\text{ch. part.}$ measured in different Pb--Pb centrality intervals~\cite{ALICE:2018rtz} at midrapidity are utilized for $v_{n}^\text{assoc}$. The measured $p_\text{T}$ spectra of muons from heavy-flavor hadron decays~\cite{ATLAS:2021xtw} and charged particles~\cite{ALICE:2018vuu} are used to compute a statistically weighted average of the $v_n$ values within the $\pt$ ranges relevant to this analysis.

The von Mises distribution approximates the circular analogue of the normal distribution and is used in directional statistics to model the distribution of periodic data~\cite{fishersphere}. For this reason, it is employed in similar analyses of pp and p--Pb collisions to characterize the NS and AS peaks~\cite{ALICE:2023kjg}. The following fit function is applied to the fully-corrected per-trigger correlation distribution, 
\begin{equation}
    f(\Delta\varphi) = B(\Delta \varphi) + \frac{Ae^{\kappa_\mathrm{NS}\mathrm{cos(\Delta\varphi)}}}{2\pi I_{0}(\kappa_\mathrm{NS})}.
\label{eqn:vonMises_function}
\end{equation}

Equation~\ref{eqn:vonMises_function} is composed of a von Mises function to describe the NS peak, combined with the baseline function $B(\Delta\varphi)$ from Eq.~\ref{eqn:Baseline_function}, which accounts for the uncorrelated background and flow contributions. For the von Mises function, the mean is fixed at $\Delta\varphi = 0$ to characterize the NS peak. The $v_{n}$ values in $B(\Delta\varphi)$ are fixed, while the other parameters are left free. The $\kappa_\mathrm{NS}$ term quantifies the concentration of the NS peak, where $1/\kappa_\mathrm{NS}$ corresponds to the variance $\sigma^{2}$ for a circular distribution. $I_{0}$ denotes the zeroth-order modified Bessel function evaluated at $\kappa_\mathrm{NS}$, and $A$ represents the integral of the NS peak. 

For some $p_\mathrm{T}^\mathrm{assoc}$ intervals in both centrality classes, no AS peak is observed. When an AS peak is evident, its shape is not consistently described by a von Mises distribution. Due to this $p_\mathrm{T}$-dependent behavior and to ensure a stable fit for determining the baseline position, the correlation distribution is fitted with Eq.~\ref{eqn:vonMises_function} in the transverse region and the NS peak region, excluding the AS peak range $1.96 < \Delta\varphi < 4.3$~rad. The baseline function position is then determined from this fit procedure for all $p_\mathrm{T}^\mathrm{assoc}$ intervals. After subtracting the baseline, the per-trigger yields are obtained by integrating the distribution within the NS and AS peak regions in $\Delta\varphi$, as detailed in Sec.~\ref{chp:results}.

\section{Systematic uncertainties}
\label{chp:systematics}
The $\Delta\varphi$ correlation distribution and the resulting NS and AS per-trigger yield observables are subject to systematic uncertainties related to the analysis procedure. These uncertainties originate from the electron-track selection process, electron identification and subtraction of the hadron contamination, the background-electron subtraction, the associated-particle efficiency correction, the mixed-event correction procedure, and the baseline estimation on the correlation distribution. Each of these uncertainty sources is estimated separately by varying the selection criteria or by introducing an alternative approach to estimate a specific quantity. The systematic uncertainty is evaluated by examining the ratio of the correlation distribution or the per-trigger yield obtained from each variation to that from the nominal configuration.
The systematic uncertainties on the correlation distribution from the associated-particle efficiency and the mixed-event correction are considered correlated since their source contribution varies by the same amount within $\Delta\varphi$, while all other sources are considered uncorrelated due to the uncertainty varying as a function of $\Delta\varphi$. 

The evaluation of each systematic uncertainty is performed separately for each trigger- and associated-$p_\text{T}$ interval, and for each centrality interval. A summary of the systematic uncertainties affecting the correlation distribution, the NS and the AS peaks, for $4 < p_\text{T}^\text{e} < 12$ GeV/$c$, are reported in Tables~\ref{tab:SysCombined_412_010_3050} and~\ref{tab:SysCombined_baseline}  for the 0--10\% and 30--50\% centrality classes. Correlated and uncorrelated uncertainties in $\Delta\varphi$ are separated in Table~\ref{tab:SysCombined_412_010_3050}. The total uncertainty is reported for the per-trigger peak yields in this table. Table~\ref{tab:SysCombined_baseline} reports the baseline pedestal value ($b$) obtained from the nominal baseline determination method (Eq.~\ref{eqn:vonMises_function}) for each $p_\mathrm{T}^\mathrm{assoc}$ interval, along with the corresponding absolute systematic uncertainties estimated from variations in the baseline determination. The NS and AS peak yields and their absolute systematic uncertainties arising solely from baseline estimation, are reported for both centrality classes in this table. 
\linebreak

\begin{table}[H]
\centering
\caption{Systematic uncertainties from various sources (excluding baseline estimation) on the azimuthal-correlation distribution, as well as NS and AS peak yields, are reported for $4 < p_\text{T}^\text{e} < 12$ GeV/$c$ in 0--10\% (30--50\%) central Pb--Pb collisions. Total uncertainties from $\Delta\varphi$-correlated and uncorrelated sources on the correlation distribution are provided separately.}
\begin{tabular}{ l c c c c c}
 \hline
 Source & $\Delta \varphi$ distribution & NS yield & AS yield \\
 \hline
 Electron-track selection & 5\% (2\%) & 5\% (2\%) & 5\% (3\%) \\
 Electron identification & 3\% (1\%) & 3\% (2\%) & 4\% (2\%) \\
 Background electron & 5\% (3\%) & 5\% (3\%) & 5\% (4\%) \\
 Associated particle selection & 5\% (3\%) & 5\% (3\%) & 5\% (3\%) \\
 Mixed-event correction & 2\% (2\%) & 2\% (2\%) & 2\% (2\%) \\ 
  \hline
 
 Total (correlated sources) & 5.4\% (3.6\%) &  & \\ 
 Total (uncorrelated sources) & $<7.7$\% ($<3.7$\%) &  & \\ 
 Total &  & 9.4\% (7.1\%) & 9.7\% (7.6\%) \\ 
  \hline
\end{tabular}
\label{tab:SysCombined_412_010_3050}
\end{table}

\begin{table}[tbp]
\centering
\caption{Estimate of the pedestal value ($b$) of the azimuthal-correlation distributions together with its absolute systematic uncertainty in the interval $4 < p_\text{T}^\text{e} < 12$ GeV/$c$ in Pb--Pb collisions in the 0–10\% and 30–50\% centrality classes. The NS and AS associated peak yields are also reported together the absolute systematic uncertainty due to the baseline estimation.}
\begin{tabular}{ l c c c c c}
 \hline
 $\pt^\mathrm{assoc}$ (\GeVc) & $b$ ($\rm{rad}^{-1}$) & NS yield & AS yield \\
 \hline
0--10\% centrality class
\\
 \hline
 $1 - 2$ & 103.0 $\pm$ 0.065  & 0.953 $\pm$ 0.154 & 0.765 $\pm$ 0.206 \\
 $2 - 3$ & 15.37 $\pm$ 0.025 & 0.412 $\pm$ 0.039 & 0.155 $\pm$ 0.078 \\
 $3 - 4$ & 2.429 $\pm$ 0.010  & 0.134 $\pm$ 0.016 & 0.035 $\pm$ 0.033 \\
 $4 - 5$ & 0.357 $\pm$ 0.003  & 0.042 $\pm$ 0.004 & 0.010 $\pm$ 0.010 \\
 $5 - 7$ & 0.061 $\pm$ 0.002  & 0.027 $\pm$ 0.002 & 0.005 $\pm$ 0.005 \\
  \hline
30--50\% centrality class \\
\hline
$1 - 2$ & 24.28 $\pm$ 0.037  & 0.608 $\pm$ 0.087 & 0.006 $\pm$ 0.115 \\ 
$2 - 3$ & 3.756 $\pm$ 0.014  & 0.277 $\pm$ 0.022 & 0.162 $\pm$ 0.043 \\
 $3 - 4$ & 0.685 $\pm$ 0.006  & 0.116 $\pm$ 0.009 & 0.023 $\pm$ 0.018 \\
 $4 - 5$ & 0.118 $\pm$ 0.003  & 0.042 $\pm$ 0.003 & 0.018 $\pm$ 0.008 \\
$5 - 7$ & 0.023 $\pm$ 0.001  & 0.027 $\pm$ 0.002 & 0.017 $\pm$ 0.004 \\ 
  \hline
\end{tabular}
\label{tab:SysCombined_baseline}
\end{table}

To estimate the uncertainty due to potential biases introduced by the electron-candidate track-quality selection, the same type of quality selection criteria to what is reported in Ref.~\cite{ALICE:2019nuy} are varied, in particular, requirements on the number of crossed TPC pad rows, the number of crossed rows over findable clusters. The number of crossed rows requirement is varied by $\pm 10$ from the nominal value, and the number of crossed rows over findable clusters is varied by $\pm 0.1$ from the nominal value. For the 0--10\% (30--50\%) centrality interval, there is an estimated~5\% (2\%) uncertainty on the correlation distribution in the trigger interval of $4 < p_\text{T}^\text{e} < 12$ GeV/$c$. The systematic uncertainty for the electron-track selection is estimated to be within 5\% (2\%) for the NS peak yields, and 5\% (3\%) for the AS peak yields in the 0--10\% (30--50\%) centrality classes. 

The uncertainty on the estimation of the hadron contamination affecting the purity of the electron sample, and on the stability of the electron-identification criteria, and their impact on the correlation distribution is estimated by varying the identification criteria on the TPC and EMCal signals, i.e., of 
$n\sigma_\text{e}^\text{TPC}$, $E/p$, and $\sigma^{2}_\text{long}$. A full background subtraction procedure is performed for each variation. The variations in the selection criteria result in a change of electron purity and efficiency of 3\% and 10\%, respectively. 
The total uncertainty determined from these sources is 3\% (1\%) for the $\Delta\varphi$ distribution, 3\% (2\%) for the NS peak yields, and 4\% (2\%) for the AS peak yields in the 0--10\% (30--50\%) centrality class.

The background-electron contribution from photon conversion and Dalitz decays is estimated via the invariant-mass method. A systematic uncertainty estimation on such a correction is performed by varying the selection of the partner electron tracks, such as the minimum $p_\textrm{T}$ of the partner electron ($\pm 0.1$ \GeVc~from the nominal value of 0.3 \GeVc, the track-quality selection~\cite{ALICE:2019nuy}, and the invariant-mass threshold of the electron-positron pairs (10 MeV/$c^{2}$ around the nominal value of 140 MeV/$c^{2}$). These variations primarily affect the tagging efficiency, which varies by 20\% (17\%) in the 0--10\% (30--50\%) centrality class. The systematic uncertainty is 5\% (3\%) for the $\Delta\varphi$ distribution, 5\% (3\%) for the NS peak yields, and 5\% (4\%) for the AS peak yields, for $4 < p_\text{T}^\text{e} < 12$ GeV/$c$ in 0--10\% (30--50\%) centrality class, respectively.

The uncertainty resulting from the track-selection criteria applied to the associated charged particles is estimated by varying the number of TPC crossed pad rows ($\pm 10$), the number of crossed rows over findable clusters ($\pm 0.1$), and the $\eta$ requirement ($\pm 0.1$) from the nominal value. This uncertainty is considered correlated in $\Delta\varphi$. Uncertainty values of 5\% (3\%) for both the distribution and peak yields are obtained for \linebreak $4 < p_\text{T}^\text{e} < 12$ GeV/$c$ interval for the 0--10\% (30--50\%) centrality class.

To estimate the uncertainty due to the mixed-event correction, the criteria used for normalizing the mixed-event distributions are varied by slightly changing the multiplicity and $z$-vertex selection pools. As an additional systematic check, the mixed-event normalization factor $\beta$ (see Eq.~\ref{eqn:Mixed_Event}) is obtained considering the integrated distribution yield over the full $\Delta\varphi$ range for $|\Delta\eta| < 0.01$. A correlated uncertainty of 2\% is estimated for the $\Delta\varphi$ distribution as well as for the NS and AS peak yields in both centrality classes in the $4<p_\text{T}^\text{e}<12$ GeV/$c$ interval.

The largest source of systematic uncertainty arises from evaluating the baseline function, which depends on the value of $b$ and the uncertainties on the measured $v_{n}$ values in Eq.~\ref{eqn:Baseline_function}. Two methods are used to estimate the systematic uncertainty of the baseline pedestal ($b$) position. First, the fit uncertainty of the $b$ parameter ($\sigma_{b}$) from Eq.~\ref{eqn:vonMises_function} is used to define two baseline variations by adjusting the $b$ parameter from its central value to $b \pm \sigma_{b}$. The second method involves fitting the correlation distribution with the baseline function, Eq.~\ref{eqn:Baseline_function}, in three transverse regions: $-\pi/2 < \Delta\varphi <-1.18$~rad (left), $|\Delta\varphi - \pi/2|< 0.4 $~rad (middle), and $4.3 < \Delta\varphi < 3\pi/2$~rad (right), excluding the NS and AS peak regions. The average $b$ parameter from these three fits is then calculated. Furthermore, the definitions of the transverse regions are varied for additional systematic checks, and the corresponding $b$ parameters are obtained.
An additional systematic check is conducted to assess whether the uncertainties in the $v_{n}$ measurement affects the baseline and peak yields. This is done by varying $v_{n}$ by $\pm \sigma_{v_{n}}$, where $\sigma_{v_{n}}$ represents the combined statistical and systematic uncertainties of the measured values.  
It is found that the uncertainty due to $\sigma_{v_{n}}$ is negligible compared to the uncertainty on the pedestal $b$. The overall systematic uncertainty is calculated by taking the difference between the largest and smallest of the considered baseline variations. 
The uncertainty from the baseline estimation on the correlation distribution is reported as an absolute value, affecting all $\dph$ bins equally. Unlike other uncertainties in the measurement, the baseline uncertainty does not intrinsically alter the overall correlation distribution but shifts the yield within a given uncertainty. Therefore, when plotting the NS and AS peak yields, the uncertainty from the baseline estimation is presented separately from other sources.
 
The above systematic estimation procedures are repeated to estimate the systematic uncertainties from the above mentioned sources on the correlation distribution, NS and AS yields for $4 < \pt^{\rm e} < 7$~\GeVc and $7 < \pt^{\rm e} < 16$~\GeVc. The uncertainty values are found to be similar to those obtained for\linebreak $4<\pt^{\rm e}<12$~\GeVc~.

\section{Results}
\label{chp:results}
The azimuthal-correlation distributions between electrons from heavy-flavor hadron decays and charged particles in 0--10\% and 30--50\% Pb--Pb collisions at $\sqrt{s_{\rm{NN}}}=5.02$ TeV are presented for different trigger electron and associated charged-particle \pt intervals. The baseline subtracted correlation distributions in Pb--Pb collisions are compared to those from pp collisions. The per-trigger associated yields $(Y_{\Delta\varphi} = N^\mathrm{assoc}/N^\mathrm{(c,b)\rightarrow e} )$ on the NS and AS of the correlation distributions are obtained by integrating the peaks in a specific ${\Delta\varphi}$ region as detailed in Sec.~\ref{sec:YieldAndIAA}, and are compared to those in pp collisions. The per-trigger yields from the two collision systems are used to calculate the per-trigger nuclear modification factor ($I_{\rm{AA}}$). The $I_{\rm{AA}}$ is defined as the ratio of the per-trigger associated yield in Pb--Pb ($Y_{\Delta\varphi}^\mathrm{Pb-Pb}$) over the per-trigger yield in pp ($Y_{\Delta\varphi}^\mathrm{pp}$)~\cite{ALICE:2022rau,ALICE:2016gso,ALICE:2011gpa}, 

\begin{equation}
I_\mathrm{AA}=\frac{Y_{\Delta\varphi}^\mathrm{Pb-Pb}}{Y_{\Delta\varphi}^\mathrm{pp}},
\label{eqn:IAA_definition}
\end{equation}

obtained in the $\Delta\varphi$ interval corresponding to the correlation peak for Pb--Pb and pp, respectively. The resulting $I_{\rm{AA}}$ are compared to those measured from correlations triggered by inclusive charged hadrons and $\rm{K}^{0}_{\rm{S}}$ mesons to study the flavor dependence of the modification of the associated-particle yields.

\begin{figure}[h!]
\centering
\includegraphics[scale=0.8]{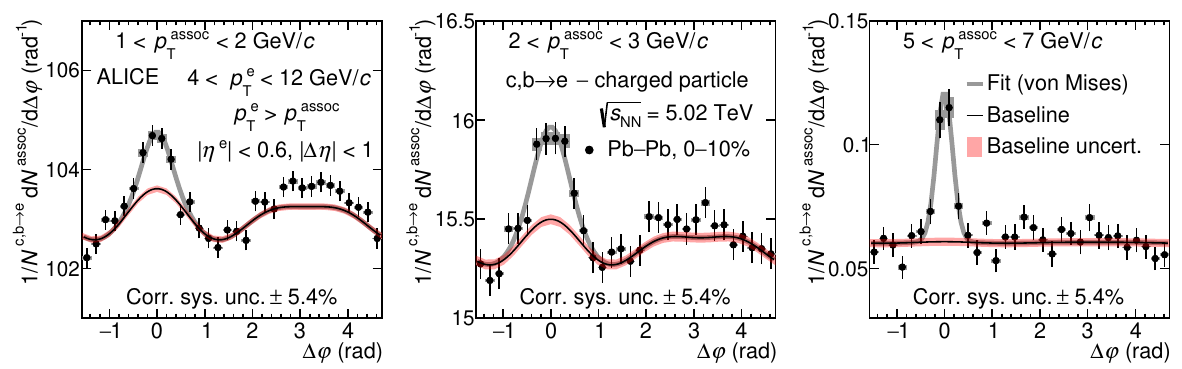}
\includegraphics[width=\linewidth]{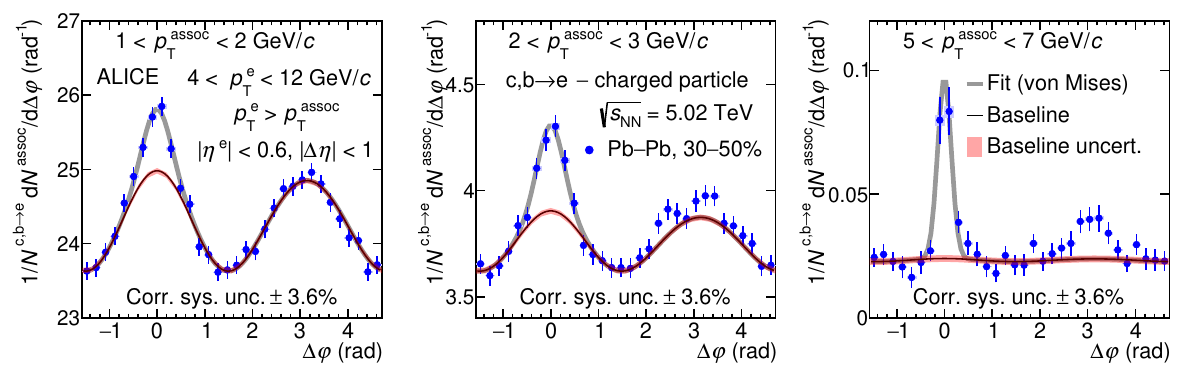}
\caption{Azimuthal-correlation distributions of electrons from heavy-flavor hadron decays and charged particles before baseline subtraction for $4 < \pt^{\rm e} < 12$~\GeVc and different associated \pt intervals, for the central (0--10\%, top panels) and semicentral (30--50\%, bottom panels) Pb--Pb collisions at $\sqrt{s_{\rm{NN}}} = 5.02$ TeV. The distributions are fitted with a von Mises function to describe the NS peak (solid gray line) and a baseline function which includes a constant term and $v_{\rm{n}}$ modulations (solid black curve). The statistical (uncorrelated systematic) uncertainties are shown as vertical lines (filled boxes). The uncertainties on the baseline are shown with a red band.}
\label{fig:DeltaPhiPbPb}
\end{figure}

\subsection{$\Delta\varphi$ distributions}
Figure~\ref{fig:DeltaPhiPbPb} shows the azimuthal-correlation distributions between trigger electrons in the interval\linebreak $4 < \pt^{\rm e} < 12$~\GeVc and associated particles in different \pt intervals, with the fit function defined in Eq.~\ref{eqn:vonMises_function}, measured in Pb--Pb collisions in 0--10\% (top panels) and 30--50\% (bottom panels) centrality intervals. The remaining $\pt$ intervals are shown in Appendix~\ref{sec:Append}.
The correlated systematic uncertainties, from the associated-particle selection and mixed-event correction, are reported in the legend for each $\pt^{\rm{assoc}}$ interval. The $\Delta\varphi$-correlated uncertainty on the baseline is shown as dotted curves. 

The baseline contribution, which depends on the charged-particle multiplicity, is higher in 0--10\% central events compared to 30--50\% central events, with its value decreasing with increasing $\pt^{\rm{assoc}}$ in both centrality intervals. A significantly large $v_{n}$ contribution dominates the AS peak at low $\pt^{\rm{assoc}}$. For both collision centrality classes, the NS peak height decreases and its width narrows as $\pt^{\rm{assoc}}$ increases.


\begin{figure}[h!]
\centering
\includegraphics[width=\linewidth]{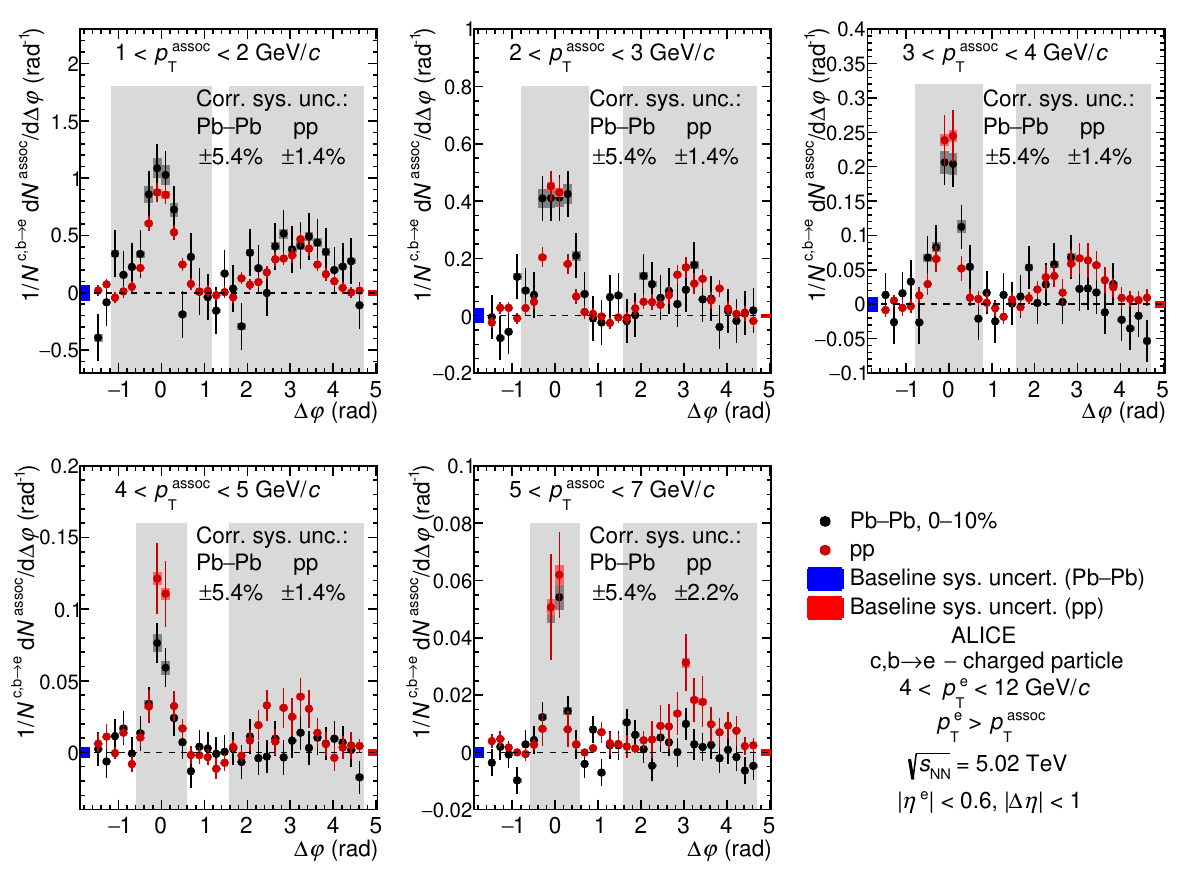}
\caption{Comparison of the azimuthal-correlation distributions of electrons from heavy-flavor hadron decays and charged particles measured in the 0--10\% centrality class Pb--Pb collisions and in pp collisions~\cite{ALICE:2023kjg},  after the baseline subtraction, for $4 < \pt^{\rm e} < 12$~\GeVc ~and different associated \pt intervals. The statistical (uncorrelated systematic) uncertainties are shown as vertical lines (filled boxes). The uncertainties on the baseline estimation are shown as solid boxes at $\dph \sim -2$ and $5$ rad. The shaded gray area corresponds to the integrated regions in $\dph$ to obtain the NS and AS yield.}
\label{fig:Delphi_CompareTopp010}
\end{figure}

\begin{figure}[h!]
\centering
\includegraphics[width=\linewidth]{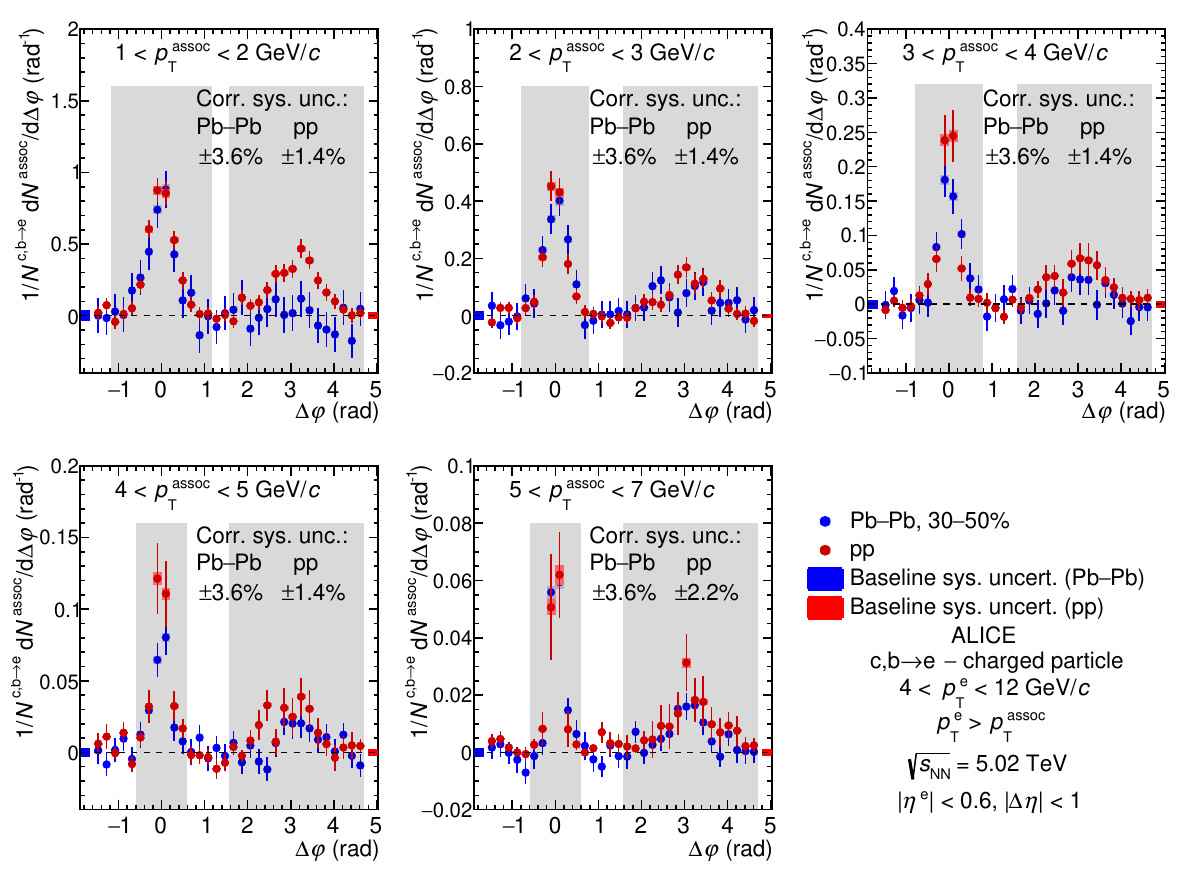}
\caption{Comparison of the azimuthal-correlation distributions of electrons from heavy-flavor hadron decays and charged particles measured in the 30--50\% centrality class Pb--Pb collisions and in pp collisions~\cite{ALICE:2023kjg},  after the baseline subtraction, for $4 < \pt^{\rm e} < 12$~\GeVc ~and different associated \pt intervals. The statistical (uncorrelated systematic) uncertainties are shown as vertical lines (filled boxes). The uncertainties on the baseline estimation are shown as solid boxes at $\dph \sim -2$ and $5$ rad. The shaded gray area corresponds to the integrated regions in $\dph$ to obtain the NS and AS yield.}
\label{fig:Delphi_CompareTopp3050}
\end{figure}

To study the impact of the QGP medium on the NS and AS peaks of the $\dph$-correlation distribution, the baseline-subtracted distributions in Pb--Pb collisions for 0--10\% and 30--50\% centrality classes are compared to those obtained in pp collisions~\cite{ALICE:2023kjg}. The comparison is performed for $4 < \pt^{\rm e} < 12$~\GeVc and for different $\pt^{\rm{assoc}}$ intervals, as shown in Fig.~\ref{fig:Delphi_CompareTopp010} for the 0--10\% centrality class and Fig.~\ref{fig:Delphi_CompareTopp3050} for the 30--50\% centrality class. 
The shape of the NS peaks for 0--10\% central Pb--Pb collisions follows a similar trend to that of pp collisions within the statistical and systematic uncertainties, such that the peak narrows as a function of $\pt^{\rm{assoc}}$. Concurrently, the AS peak tends towards a suppression compared to pp collisions for $\pt^{\rm{assoc}} > 3$~\GeVc. 
For Pb--Pb collisions in the 30--50\% centrality class, the NS peak shapes are consistent with those of pp collisions for all $\pt^{\rm{assoc}}$. However, the AS peaks in semicentral Pb--Pb collisions tend to be lower than in pp collisions within large statistical uncertainties. The differences between the peaks are further quantified by the $I_\mathrm{AA}$, as discussed below. The observed modifications on the AS of the $\dph$ distributions in Pb--Pb collisions potentially indicate the presence of interactions between the medium and the parton recoiling from the heavy quark, as well as the related shower~\cite{Vitev:2005yg, PHENIX:2008osq,PHENIX:2010cfl,ALICE:2011gpa}.

\subsection{Per-trigger associated-particle yield and per-trigger nuclear modification factor} \label{sec:YieldAndIAA}

\begin{figure}[!h]
\centering
\subfigure{
\includegraphics[width=0.48\linewidth]{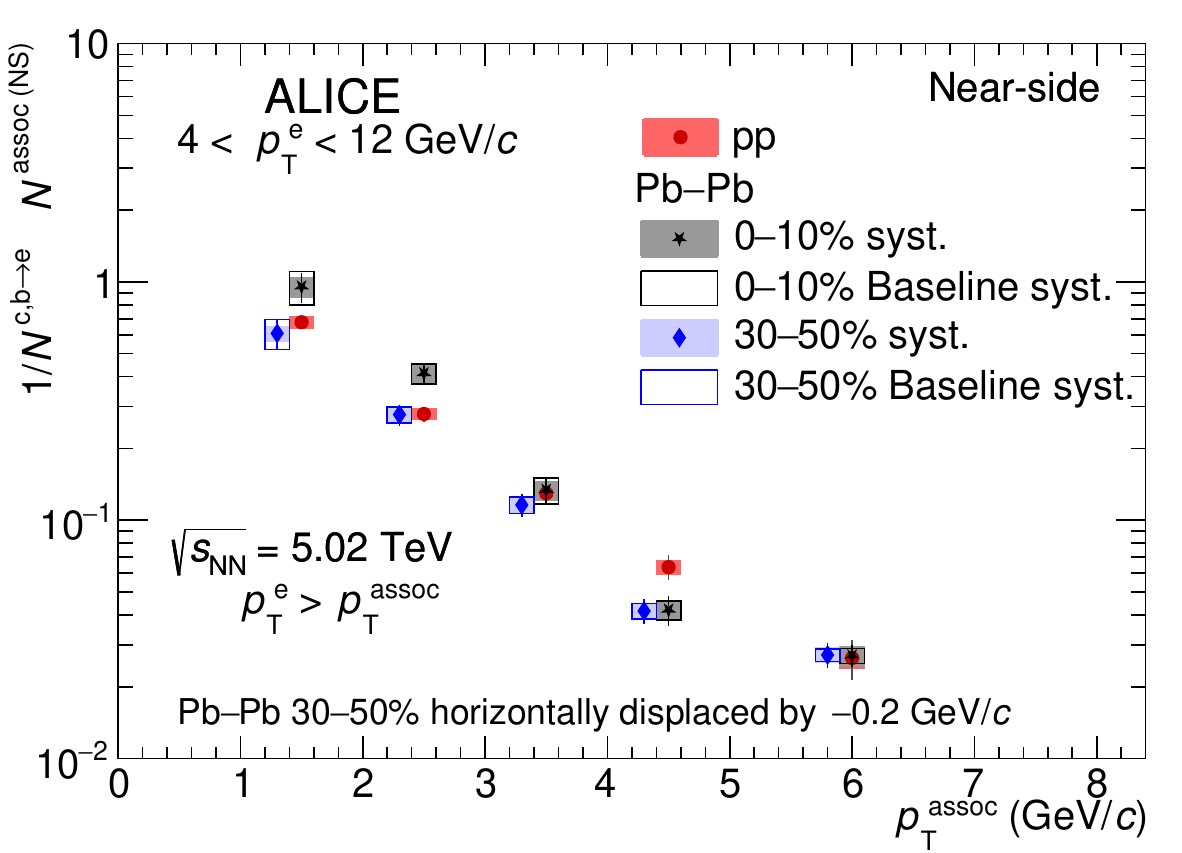}
}
\subfigure{
\includegraphics[width=0.48\linewidth]{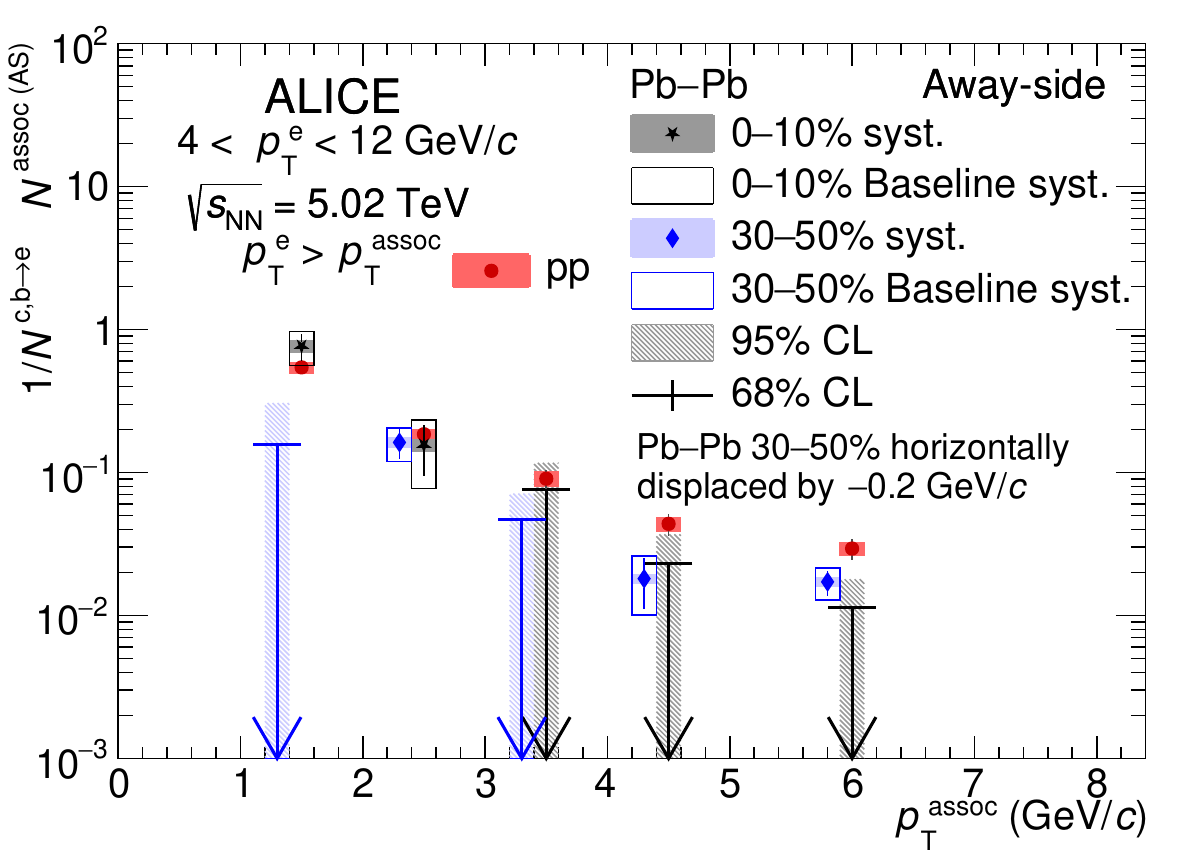}
}
\caption{Per-trigger associated yields of the NS (left) and AS (right) peaks for $4 <\pt^{\rm{e}}< 12$~\GeVc as a function of $\pt^{\rm{assoc}}$ in Pb--Pb collision centrality classes 0--10\% and 30--50\%, compared to those obtained from minimum-bias pp collisions. The statistical (systematic) uncertainties are shown as vertical lines (empty boxes). In some $\pt^{\rm{assoc}}$ intervals, the AS yield is consistent with zero within one standard deviation of statistical and systematic uncertainties added in quadrature. For those intervals, upper limits on the yields for 68\% and 95\% confidence levels are evaluated, and are shown with arrows and boxes, respectively.}
\label{fig:Final_Yields_4to12}
\end{figure}

To perform a quantitative comparison of the correlation peaks between pp and Pb--Pb collisions, the per-trigger associated yields on the NS and AS peaks are evaluated as a function of $\pt^{\rm{assoc}}$ for \linebreak $4<\pt^{\rm{e}}<12$~\GeVc, as shown in Fig.~\ref{fig:Final_Yields_4to12} for both collision centrality classes. The yields are obtained by integrating the baseline-subtracted correlation distributions in the $\pm3\sigma$ region from the center of the NS and AS peak regions along $\Delta\varphi$, where $\sigma$ is taken from the pp peak width value estimated from PYTHIA and obtained from the study in Ref.~\cite{ALICE:2023kjg}. The $\sigma_
\mathrm{NS}$ values used in this analysis (as a function of increasing $\pt^{\rm{assoc}}$) are: $\sigma_\mathrm{NS} = 0.33, 0.25, 0.2, 0.175, 0.15$. Due to the size of the uncertainty in the pp AS width measurement, the AS per-trigger yields were determined by integrating the correlation distribution in a fixed region of $\pm3\sigma_\mathrm{AS}$ (with $\sigma_\mathrm{AS}=0.5$). The integration regions for the peak yields are shown in Fig.~\ref{fig:Delphi_CompareTopp010} and Fig.~\ref{fig:Delphi_CompareTopp3050}. This integration region is the same for distributions from both Pb--Pb and pp collisions. The systematic uncertainty on the yield due to the baseline estimation is presented separately (empty box) from other sources (solid box). 

The NS and AS yields are observed to decrease with increasing $\pt^{\rm{assoc}}$ in pp, p--Pb~\cite{ALICE:2023kjg}, and Pb--Pb collision systems. This trend can be understood in terms of the relatively large momentum fraction carried by the heavy-flavor hadron in the fragmentation process, which leaves less energy available for additional particles and thus leads to associated particles being produced predominantly at low $p_{\mathrm{T}}$.  The AS Pb--Pb yield in several $\pt^{\rm{assoc}}$ intervals is consistent with zero within one standard deviation, accounting for statistical and systematic uncertainties combined in quadrature. This may be due to a strong suppression from in-medium energy loss. In these intervals, upper limits of the yield value at 68\% and 95\% confidence levels are evaluated and are represented with arrows. 

\begin{figure}[!h]
\centering
\subfigure{
\includegraphics[scale=0.38]{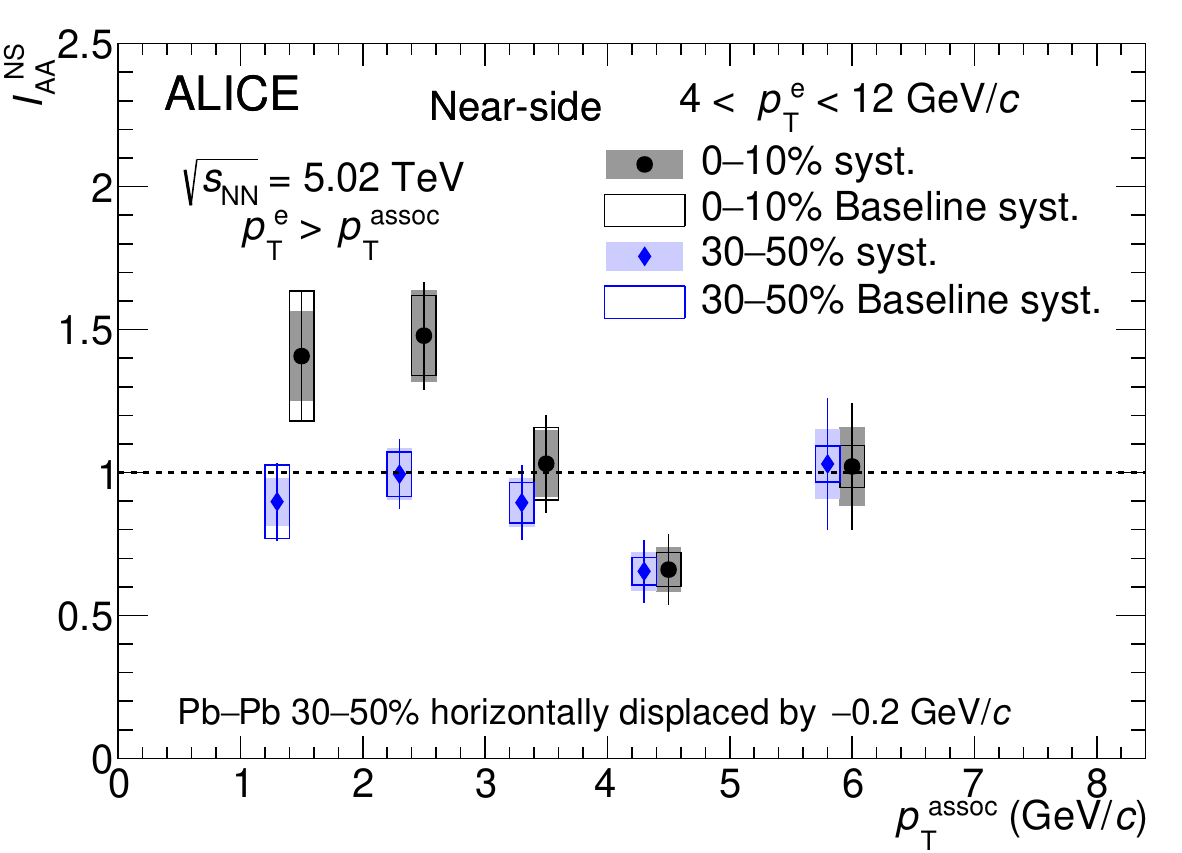}
}
\subfigure{
\includegraphics[scale=0.38]{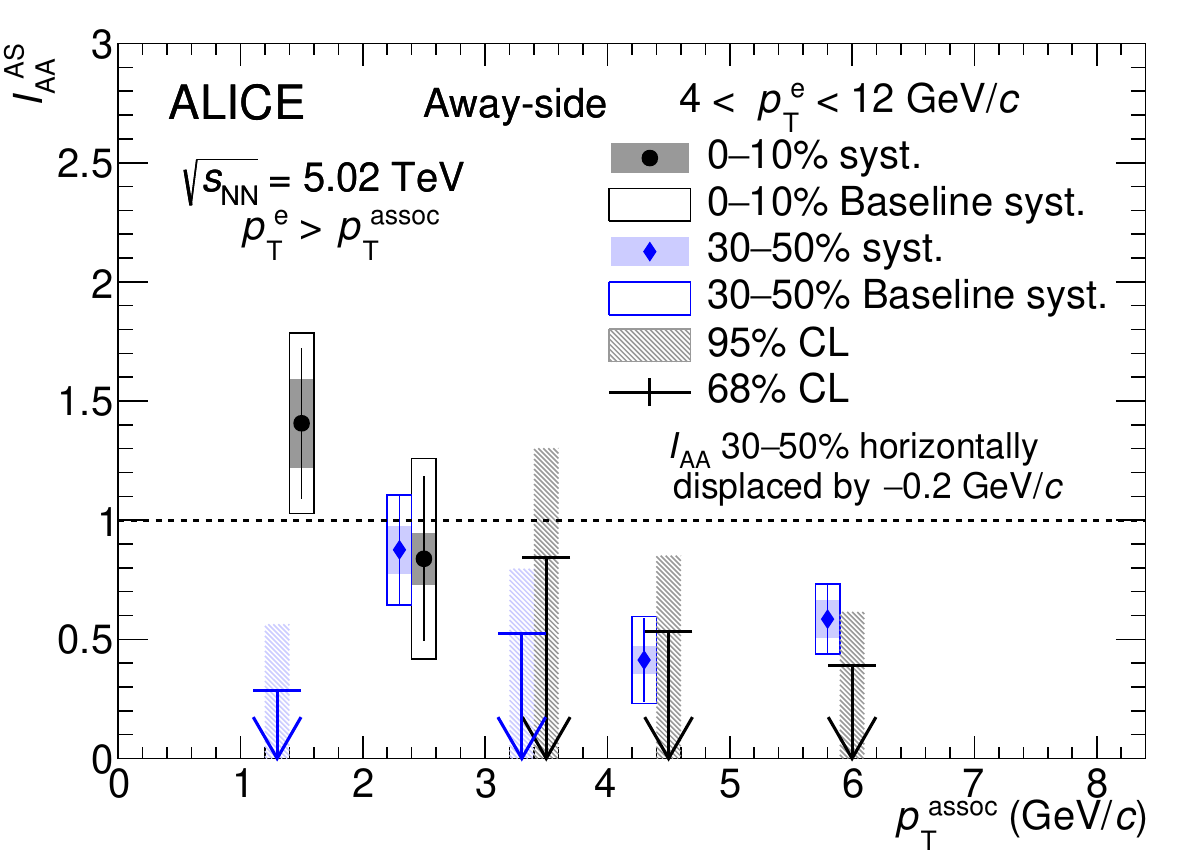}
}
\caption{Per-trigger nuclear modification factor ($I_{\rm{AA}}$) of NS (left) and AS (right) associated yields, for electrons from heavy-flavor hadron decays in the interval $4 < \pt^{\rm{e}}< 12$~\GeVc as a function of $\pt^{\rm{assoc}}$ in Pb--Pb collisions of centrality classes 0--10\% and 30--50\%. The statistical (systematic) uncertainties are shown as vertical lines (empty boxes). In some $\pt^{\rm{assoc}}$ intervals the AS yield is consistent with zero within one standard deviation of total uncertainty. For those intervals, upper limits on the $I_{\rm{AA}}$ for 68\% and 95\% confidence levels are shown with arrows and boxes, respectively.}
\label{fig:Final_IAA_4to12}
\end{figure}

The ratios of the per-trigger associated NS and AS yields in Pb--Pb to pp collisions are calculated in each kinematic interval to obtain the per-trigger nuclear modification factors, $I_{\rm{AA}}$. The $I_{\rm{AA}}$ observable is sensitive to changes in the heavy-quark fragmentation function and thus provides a quantifiable measure of the QGP medium's influence on the jet relative to the vacuum. The modification of the jet which contains the heavy quark is described by the NS $I_{\rm{AA}}$, whereas the modification of the recoil jet is described by the AS $I_{\rm{AA}}$. The $I_{\rm{AA}}$ of associated yields on the NS and AS for both collision centrality classes are shown as a function of $\pt^{\rm{assoc}}$ for $4 < \pt^{\rm{e}} < 12$~\GeVc~in Fig.~\ref{fig:Final_IAA_4to12}. For the kinematic intervals showing AS yields consistent with zero within one standard deviation of total uncertainties, the related $I_{\rm{AA}}$ are also reported as upper limits for 68\% and 95\% confidence levels, and shown with an arrow. 

In 0--10\% central Pb--Pb collisions, the $I_{\rm{AA}}$ on the NS shows a slight hint of an increase at $\pt^{\rm{assoc}} < 3~\GeVc$, albeit with a significance of only 1.27$\sigma$ for $1 < \pt^{\rm{assoc}} < 3~\GeVc$, considering the statistical, systematic, and baseline uncertainties combined in quadrature. There are multiple possible interplaying mechanisms that would cause an enhancement, such as $k_\mathrm{T}$ broadening, medium excitation, or possible fragments from radiated gluons~\cite{Vitev:2002pf, Kopeliovich:2002yh, Wang:1998ww, Ma:2010dv, Vitev:2005yg}. At higher $\pt^{\rm{assoc}}$, the $I_{\rm{AA}}$ is consistent with unity. On the AS, the $I_{\rm{AA}}$ shows a $\pt^{\rm{assoc}}$ dependence, with values consistent with unity at low $\pt^{\rm{assoc}}$. These values decrease to approximately 0.5 for $\pt^{\rm{assoc}} > 4$~\GeVc, indicating a suppression in the higher $\pt$ interval, with a significance of 2.5$\sigma$ below unity for $\pt > 4$~\GeVc. The larger suppression of the AS $I_{\rm AA}$ relative to the NS $I_{\rm AA}$ is expected due to a trigger bias in the selection. Electrons that satisfy the trigger requirements predominantly originate from quarks produced near the surface of the QGP, which are subject to small quenching effects. Consequently, the AS parton traverses, on average, a larger path length through the medium than the NS parton, and therefore experiences a greater energy loss. The suppression of the away-side $I_{\rm{AA}}$ follows a similar behavior to what is observed in $\mathrm{D}^{0}$-tagged jets measured in Pb--Pb collisions~\cite{ALICE:2024gqr}, where the $R_{\rm{AA}}$ of the $\mathrm{D}^{0}$ jet displays a strong suppression of the jet production, indicative of the medium-induced energy loss reducing the production of jet particles. In semicentral Pb--Pb collisions, the NS $I_{\rm{AA}}$ is consistent with unity for all $\pt^{\rm{assoc}}$ intervals of the measurement while the AS $I_{\rm{AA}}$ is compatible with unity for $\pt^{\rm{assoc}} < 3$ GeV/$c$ and shows values lower than unity for $\pt^{\rm{assoc}} > 3$ GeV/$c$, albeit with large uncertainties. 

\subsection{Dependence of the correlation distributions on the trigger-electron $\pt$} \label{sec:pTeBinSplitStudies}
In pp and p--Pb collisions, the azimuthal-correlation distributions were also measured for two independent trigger electron \pt intervals, $4 < \pt^{\rm e} < 7$~\GeVc ~and $7 < \pt^{\rm e} < 16$~\GeVc,  to study the dependence on the quark flavor (charm and beauty) from which the trigger electron originates~\cite{ALICE:2023kjg}. The fractions of electrons produced by charm- and beauty-hadron decays have a strong \pt dependence, and the beauty contribution is expected to dominate in the higher \pt interval~\cite{ALICE:2014aev,Cacciari:2012ny}. In pp collisions the electrons from beauty-hadron decays at $\pt^{\rm e} = 4~\GeVc$ account for about 40\% of the total electron yield, increasing to 60–70\% for $\pt^{\rm e} > 8~\GeVc$. It is observed that the per-trigger associated yield is larger for the higher $\pt^{\rm{e}}$ interval, by about 30\% at low $\pt^{\rm{assoc}}$ and by about a factor of 10 in the highest $\pt^{\rm{assoc}}$ interval, for both pp and p--Pb collisions~\cite{ALICE:2023kjg}. By comparing the measurement to PYTHIA Monte Carlo simulations, this behavior is understood to result from heavy quarks of higher energy producing higher-\pt electrons in tandem with a greater number of associated fragmentation particles.

\begin{figure}[!h]
\centering
\subfigure{
\includegraphics[width=0.48\linewidth]{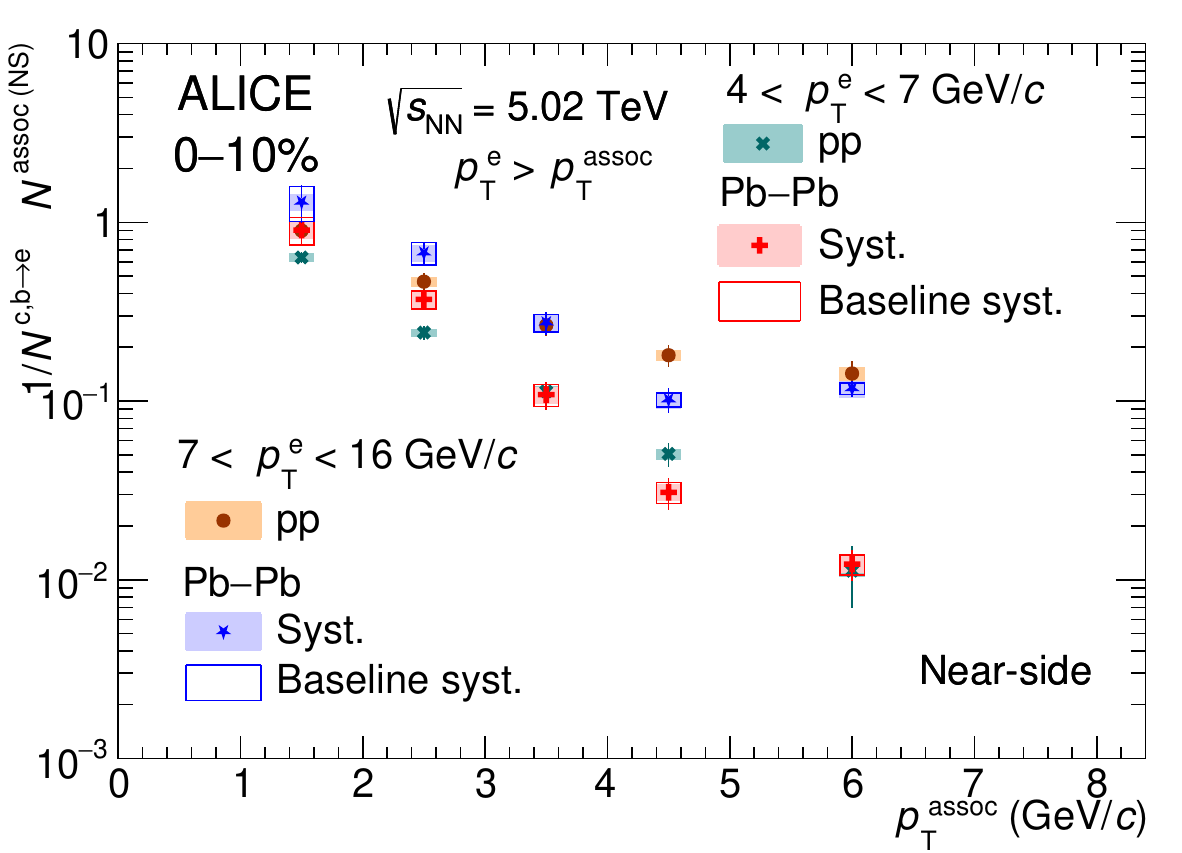}
}
\subfigure{
\includegraphics[width=0.48\linewidth]{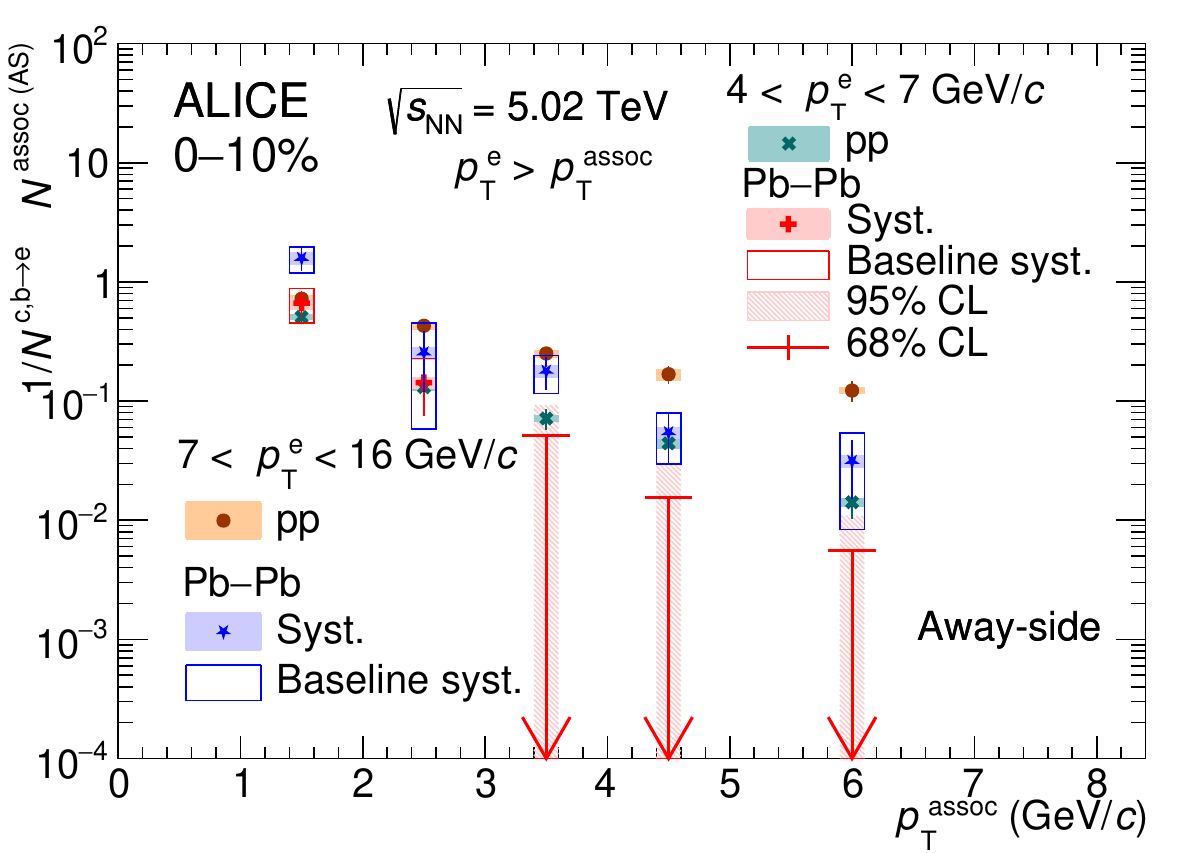}
}
\caption{Per-trigger NS (left) and AS (right) associated peak yields for electrons from heavy-flavor hadron decays in the intervals $4 < \pte < 7$~\GeVc and $7 < \pte < 16$~\GeVc as a function of $\ptassoc$ in central Pb--Pb collisions, compared to measurements in pp collisions~\cite{ALICE:2023kjg}. The statistical (systematic) uncertainties are shown as vertical lines (empty boxes). In some $\ptassoc$ intervals the AS yield is consistent with zero within one standard deviation of total uncertainties. For those intervals, upper limits on the yields for 68\% and 95\% confidence levels are shown with arrows and boxes, respectively.}
\label{fig:Final_Yields_pTSplit}
\end{figure}

In Pb--Pb collisions, the presence of a QGP induces partonic energy loss, as indicated by the nuclear modification factor ($R_{\rm{AA}}$), which depends on the parton momentum and mass. The $R_{\rm{AA}}$ of electrons from heavy-flavor hadron decays~\cite{ALICE:2019nuy} is approximately $\sim 0.35$ within the $\pt^{\rm e}$ interval of $4-16$~\GeVc, with a trend indicating an increasing $R_{\rm{AA}}$ value towards higher $\pt^{\rm e}$. The $R_{\rm{AA}}$ of particles from beauty-hadron decays is observed to be larger than that of charm hadrons~\cite{ALICE:2022tji,ALICE:2022iba, ATLAS:2021xtw}. 
The dependence of the modification of the per-trigger associated-particle production on partonic mass and initial parton energy in the presence of QGP can be studied by measuring the correlation distributions of electrons from heavy-flavor hadron decays and charged particles, and their peak yields for different $\pt^{\rm e}$ intervals. Figure~\ref{fig:Final_Yields_pTSplit} shows the comparisons of the NS and AS yields for the two $\pt^{\rm e}$ intervals $4 < \pt^{\rm e} < 7$~\GeVc ~and $7 < \pt^{\rm e} < 16$~\GeVc in 0--10\% central Pb--Pb collisions, compared with those obtained for pp collisions. The per-trigger NS and AS yields are systematically higher for the $7 < \pt^{\rm e} < 16$~\GeVc ~interval compared to the values obtained for $4 < \pt^{\rm e} < 7$~\GeVc, similar to the observations in pp and p--Pb collisions~\cite{ALICE:2023kjg}. This is expected in Pb--Pb collisions as well, as more energetic heavy quarks produce higher-\pt electrons and a greater number of accompanying particles from the quark fragmentation.

\begin{figure}[!h]
\centering
\subfigure{
\includegraphics[width=0.48\linewidth]{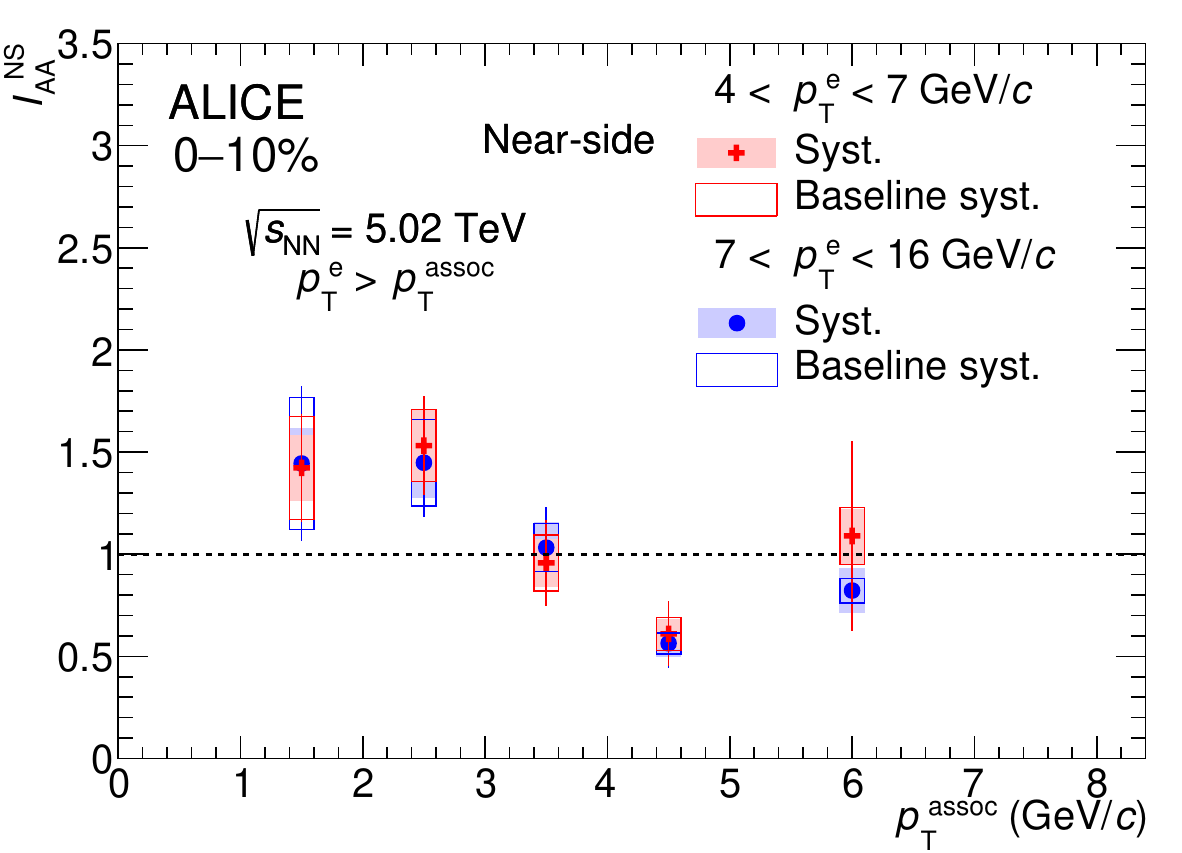}
}
\subfigure{
\includegraphics[width=0.48\linewidth]{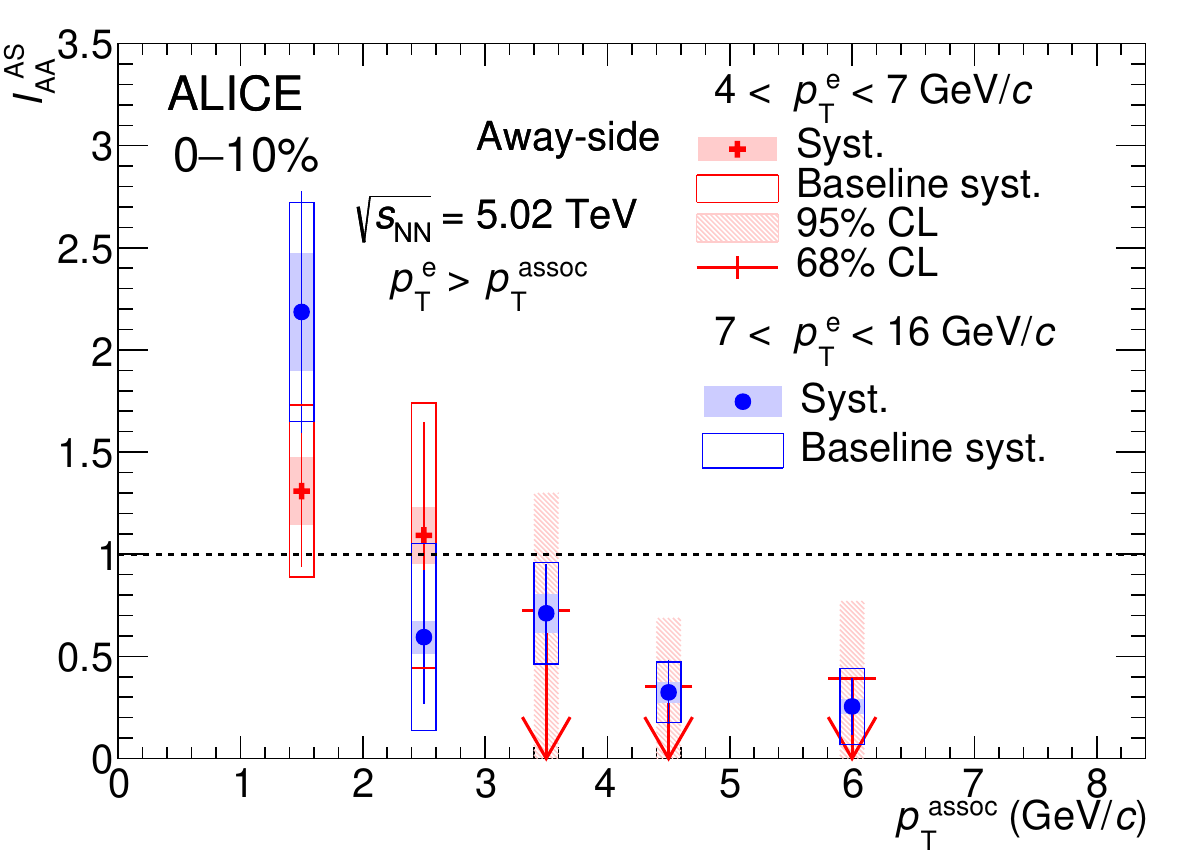}
}
\caption{Per-trigger nuclear modification factor ($I_{\rm{AA}}$) of the NS (left) and AS (right) associated peak yields for electrons from heavy-flavor hadron decays in intervals $4 < \pt^{\rm{e}} < 7$~\GeVc compared to $7 < \pt^{\rm{e}} < 16$~\GeVc as a function of $\pt^{\rm{assoc}}$ in central Pb--Pb collisions. The statistical (systematic) uncertainties are shown as vertical lines (empty boxes). In some $\pt^{\rm{assoc}}$ intervals the AS yield is consistent with zero within one standard deviation of total uncertainty. For those intervals, upper limits on the $I_{\rm{AA}}$ for 68\% and 95\% confidence levels are shown with arrows and boxes, respectively.}
\label{fig:Final_IAA_pTSplit}
\end{figure}

Figure~\ref{fig:Final_IAA_pTSplit} shows the corresponding per-trigger nuclear modification factors $I_{\rm AA}$ for the two $\pt^{\rm e}$ intervals. The AS $I_{\rm AA}$ for the $7 < \pt^{\rm e} < 16$~\GeVc interval exhibits smaller uncertainties compared to the $4 < \pt^{\rm e} < 7$~\GeVc interval and shows a suppression at $>4$~\gevc with a significance of $2.7\sigma$. Overall, the $I_{\rm AA}$ values for the two $\pt^{\rm e}$ intervals are consistent within statistical and systematic uncertainties on both the NS and AS. This indicates that, within the current uncertainties, the modification of the parton shower producing the associated-particle yield does not significantly depend on the energy and mass of the jet-initiating parton.

\subsection{Comparison of the per-trigger nuclear modification factor to that of light-flavor trigger particles} \label{sec:CompareToLF}

The comparison of the $R_{\mathrm{AA}}$ of light-flavor, charm, and beauty hadrons shows a mass-dependent ordering in the low- and intermediate-$\pt$ regions~\cite{ALICE:2021rxa,ALICE:2022tji}. Pions have the lowest $R_{\rm{AA}}$, followed by charm hadrons, with beauty hadrons exhibiting the largest $R_{\rm{AA}}$ values. This indicates a  mass-dependent partonic interaction in the QGP. Comparing the $I_{\rm{AA}}$ of the correlation peak yields from trigger particles with different light-flavor quark content provides insight into the initiating-parton mass and color-charge dependence of the QGP-induced modification of associated-particle yields. 

\begin{figure}[!h]
\centering
\subfigure{
\includegraphics[width=0.48\linewidth]{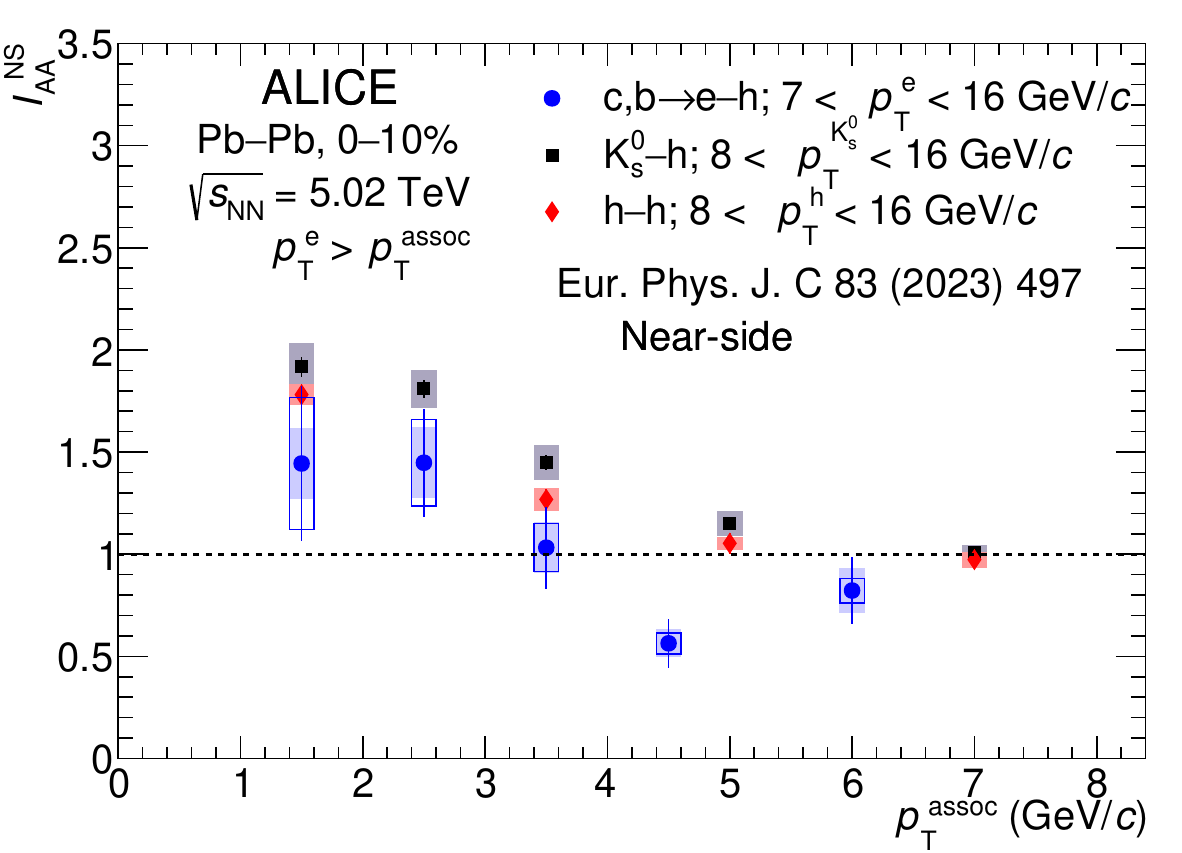}
}
\subfigure{
\includegraphics[width=0.48\linewidth]{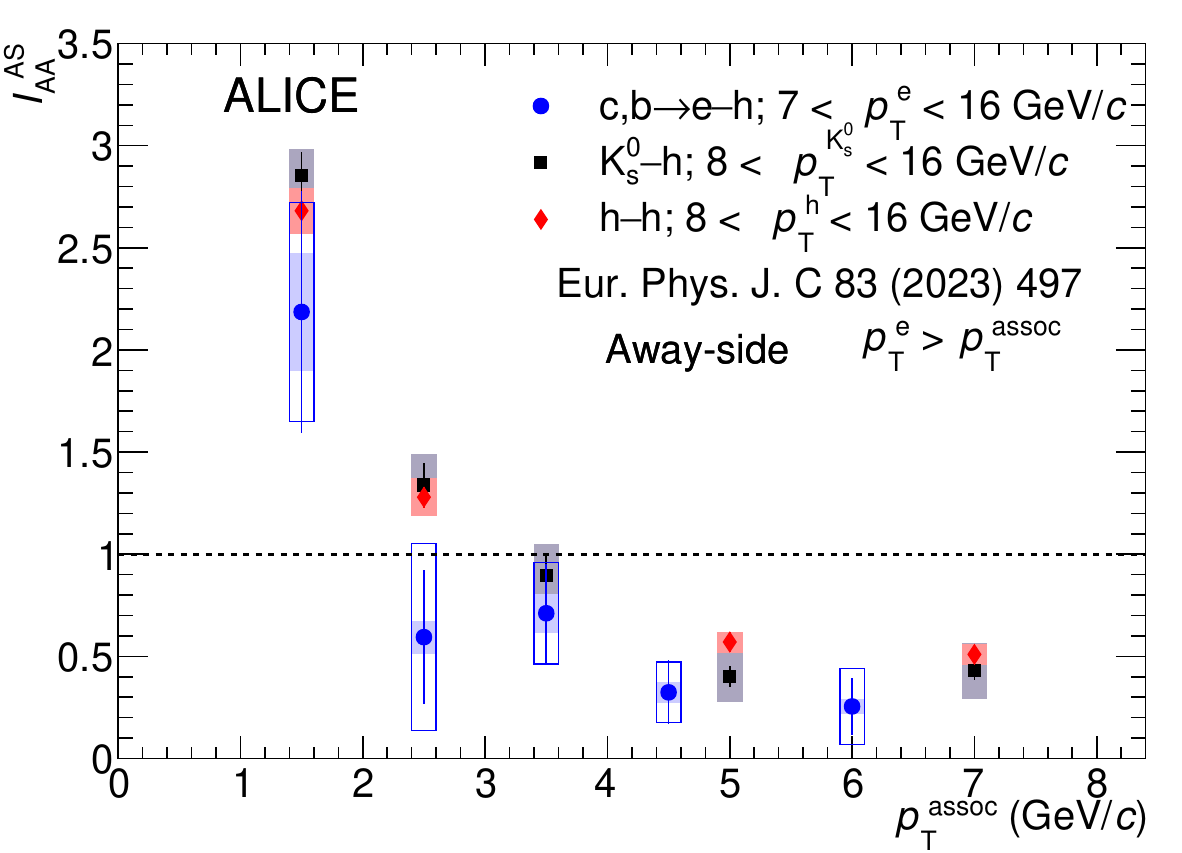}
}
\caption{$I_{\rm{AA}}$ of NS (left) and AS (right) associated peak yield from correlation distributions of electron from heavy-flavor hadron decays as the trigger particle ($7 < p_\text{T}^\text{e} < 16$~\GeVc) compared with that of charged hadrons and $\mathrm{K}^{0}_\mathrm{S}$ as trigger particles ($8 < \pt^{\rm{trig}} < 16$~\GeVc~\cite{ALICE:2022rau}).}
\label{fig:Final_IAA_DiffFlav}
\end{figure}

The NS and AS $I_{\rm{AA}}$ of associated peak yields from angular-correlation distributions of electrons from heavy-flavor hadron decays and charged hadrons are compared with those of dihadron~\cite{ALICE:2022rau}, and $\mathrm{K}^{0}_{\mathrm{S}}$ charged-hadron correlations~\cite{ALICE:2022rau}, as shown in Fig.~\ref{fig:Final_IAA_DiffFlav}. The trigger $\pt$ intervals used for the comparison are $7 < \pt^\mathrm{e} < 16$~\GeVc for electrons (where electrons from beauty-hadron decays dominate), $8 < \pt^\mathrm{trig} < 16$~\GeVc for dihadron (with a large component of charged hadrons being pions) and $\mathrm{K}^{0}_{\mathrm{S}}$-hadron correlations. The $I_{\rm{AA}}$ of dihadron and $\mathrm{K}^{0}_{\mathrm{S}}$ charged-hadron correlations shows an enhancement of low-$\pt$ associated particles on the NS, and $I_{\rm{AA}}$ consistent with unity observed for high-$\pt$ associated particles. On the AS, the associated charged-hadron yield in Pb--Pb collisions is suppressed relative to pp collisions at $\pt^\mathrm{assoc} > 4$ \gevc, but an enhancement increasing with decreasing transverse momentum is observed at low $\pt^{\rm{assoc}}$~\cite{ALICE:2022rau}. The $I_{\rm{AA}}$ values of NS and AS yields for electron triggers from heavy-flavor hadron decays are consistent with those for light-flavor and strange-particle triggers given the current uncertainties. This indicates no mass dependence on the modification of the associated yield. However, it should be noted that the parton-to-hadron $\pt$ scaling differs between heavy-flavor and light-flavor quarks, thus the corresponding parton \pt intervals may not be the same. Additionally, the kinematic differences between the electrons from heavy-flavor hadron decays and $\mathrm{K}^{0}_{\mathrm{S}}$ or charged hadrons -- arising from the decay kinematics of heavy-flavor hadrons -- should also be taken into account, along with the additional tracks from these decays that contribute to the NS peak.

\section{Summary}
\label{chp:summary}
Jet-like azimuthal correlations between electrons from heavy-flavor hadron decays and charged particles in Pb--Pb collisions at $\sqrt{s_{\rm{NN}}} = 5.02$ TeV are measured for the 0–10\% and 30–50\% centrality classes. Correlation distributions are obtained for trigger electrons with $4 < p_\mathrm{T}^\mathrm{e} < 12$ GeV/$c$ and associated charged particles in several $p_\mathrm{T}$ intervals from 1 to 7 GeV/$c$. The results are compared to pp collisions~\cite{ALICE:2023kjg}, and the per-trigger near-side (NS) and away-side (AS) peak yields are used to compute the per-trigger nuclear modification factor ($I_\mathrm{AA}$), which can provide insight into heavy-quark transport in the QGP.

The per-trigger yields are obtained for trigger electrons in $4 < p_\mathrm{T}^\mathrm{e} < 7$ GeV/$c$ and $7 < p_\mathrm{T}^\mathrm{e} < 16$ GeV/$c$ in 0–10\% central collisions in order to investigate the parton mass and jet-energy dependence. Similar to pp results, trigger electrons with larger $p_\mathrm{T}^\mathrm{e}$ yield more associated particles in all the $p_\mathrm{T}^\mathrm{assoc}$ intervals of the measurement in both the NS and AS peaks. The NS and AS $I_\mathrm{AA}$ values overlap within uncertainties for both $p_\mathrm{T}^\mathrm{e}$ intervals, suggesting only a weak dependence on jet energy and the initiating parton mass in this kinematic regime.

In 0–10\% central collisions, the NS $I_\mathrm{AA}$ suggests a possible enhancement at low $p_\mathrm{T}^\mathrm{assoc}$, though with limited significance (1.27$\sigma$), possibly indicating medium excitation or additional gluon radiation by the propagating high-$\pt$ parton in the QGP. At $p_\mathrm{T}^\mathrm{assoc} > 3$ GeV/$c$, the NS $I_\mathrm{AA}$ is consistent with unity. The AS peak $I_\mathrm{AA}$ is consistent with unity at low $p_\mathrm{T}^\mathrm{assoc}$ but shows a suppression with a $2.5\sigma$ significance at higher $p_\mathrm{T}^\mathrm{assoc}$, likely due to path-length-dependent jet energy loss and medium-induced parton shower modifications.

The $I_\mathrm{AA}$ of charged particles correlated with electrons from heavy-flavor hadron decays for $7 < p_\mathrm{T}^\mathrm{e} < 16$~GeV/$c$ is also compared with dihadron and $\mathrm{K}^{0}_\mathrm{S}$ correlations~\cite{ALICE:2022rau}. The $I_\mathrm{AA}$ for electron-triggered correlations is consistent, within uncertainties, with the dihadron and $\mathrm{K}^{0}_\mathrm{S}$ results at low associated transverse momentum, $p_\mathrm{T}^\mathrm{assoc}<3$ \gevc, suggesting a similar low-$p_\mathrm{T}$ enhancement. All measurements show an AS suppression at high $p_\mathrm{T}^\mathrm{assoc}$~($>4$ \gevc). The consistency of $I_\mathrm{AA}$ across different triggers suggests no clear quark mass dependence given the uncertainties, though variations in parton-to-hadron $p_\mathrm{T}$ scaling and the additional decay step for the electron from heavy-flavor hadrons could introduce kinematic differences.

These results indicate medium-induced modifications to heavy-flavor jet fragmentation in Pb--Pb collisions, with indications of an increased yield of low-\pt associated particles on the near-side compared to pp collisions. The away-side suppression of high-\pt particles indicates parton energy loss from the recoiling jet to the heavy-flavor jet. Theoretical modeling could further clarify the mechanisms governing heavy-quark propagation in the QGP. 
During the LHC Run 3, the upgraded ALICE detector~\cite{ALICE:2023udb} is estimated to collect more than an order of magnitude larger Pb--Pb data samples compared to Run 2, which is expected to reduce uncertainties. This will also allow correlation measurements with D mesons and other charm hadron triggers, which will enable a more precise understanding of the heavy-flavor jet-medium interaction in high-energy heavy-ion collisions.


\newenvironment{acknowledgement}{\relax}{\relax}
\begin{acknowledgement}
\section*{Acknowledgements}

The ALICE Collaboration would like to thank all its engineers and technicians for their invaluable contributions to the construction of the experiment and the CERN accelerator teams for the outstanding performance of the LHC complex.
The ALICE Collaboration gratefully acknowledges the resources and support provided by all Grid centres and the Worldwide LHC Computing Grid (WLCG) collaboration.
The ALICE Collaboration acknowledges the following funding agencies for their support in building and running the ALICE detector:
A. I. Alikhanyan National Science Laboratory (Yerevan Physics Institute) Foundation (ANSL), State Committee of Science and World Federation of Scientists (WFS), Armenia;
Austrian Academy of Sciences, Austrian Science Fund (FWF): [M 2467-N36] and Nationalstiftung f\"{u}r Forschung, Technologie und Entwicklung, Austria;
Ministry of Communications and High Technologies, National Nuclear Research Center, Azerbaijan;
Rede Nacional de Física de Altas Energias (Renafae), Financiadora de Estudos e Projetos (Finep), Funda\c{c}\~{a}o de Amparo \`{a} Pesquisa do Estado de S\~{a}o Paulo (FAPESP) and The Sao Paulo Research Foundation  (FAPESP), Brazil;
Bulgarian Ministry of Education and Science, within the National Roadmap for Research Infrastructures 2020-2027 (object CERN), Bulgaria;
Ministry of Education of China (MOEC) , Ministry of Science \& Technology of China (MSTC) and National Natural Science Foundation of China (NSFC), China;
Ministry of Science and Education and Croatian Science Foundation, Croatia;
Centro de Aplicaciones Tecnol\'{o}gicas y Desarrollo Nuclear (CEADEN), Cubaenerg\'{\i}a, Cuba;
Ministry of Education, Youth and Sports of the Czech Republic, Czech Republic;
The Danish Council for Independent Research | Natural Sciences, the VILLUM FONDEN and Danish National Research Foundation (DNRF), Denmark;
Helsinki Institute of Physics (HIP), Finland;
Commissariat \`{a} l'Energie Atomique (CEA) and Institut National de Physique Nucl\'{e}aire et de Physique des Particules (IN2P3) and Centre National de la Recherche Scientifique (CNRS), France;
Bundesministerium  f\"{u}r Forschung, Technologie und Raumfahrt (BMFTR) and GSI Helmholtzzentrum f\"{u}r Schwerionenforschung GmbH, Germany;
General Secretariat for Research and Technology, Ministry of Education, Research and Religions, Greece;
National Research, Development and Innovation Office, Hungary;
Department of Atomic Energy Government of India (DAE), Department of Science and Technology, Government of India (DST), University Grants Commission, Government of India (UGC) and Council of Scientific and Industrial Research (CSIR), India;
National Research and Innovation Agency - BRIN, Indonesia;
Istituto Nazionale di Fisica Nucleare (INFN), Italy;
Japanese Ministry of Education, Culture, Sports, Science and Technology (MEXT) and Japan Society for the Promotion of Science (JSPS) KAKENHI, Japan;
Consejo Nacional de Ciencia (CONACYT) y Tecnolog\'{i}a, through Fondo de Cooperaci\'{o}n Internacional en Ciencia y Tecnolog\'{i}a (FONCICYT) and Direcci\'{o}n General de Asuntos del Personal Academico (DGAPA), Mexico;
Nederlandse Organisatie voor Wetenschappelijk Onderzoek (NWO), Netherlands;
The Research Council of Norway, Norway;
Pontificia Universidad Cat\'{o}lica del Per\'{u}, Peru;
Ministry of Science and Higher Education, National Science Centre and WUT ID-UB, Poland;
Korea Institute of Science and Technology Information and National Research Foundation of Korea (NRF), Republic of Korea;
Ministry of Education and Scientific Research, Institute of Atomic Physics, Ministry of Research and Innovation and Institute of Atomic Physics and Universitatea Nationala de Stiinta si Tehnologie Politehnica Bucuresti, Romania;
Ministerstvo skolstva, vyskumu, vyvoja a mladeze SR, Slovakia;
National Research Foundation of South Africa, South Africa;
Swedish Research Council (VR) and Knut \& Alice Wallenberg Foundation (KAW), Sweden;
European Organization for Nuclear Research, Switzerland;
Suranaree University of Technology (SUT), National Science and Technology Development Agency (NSTDA) and National Science, Research and Innovation Fund (NSRF via PMU-B B05F650021), Thailand;
Turkish Energy, Nuclear and Mineral Research Agency (TENMAK), Turkey;
National Academy of  Sciences of Ukraine, Ukraine;
Science and Technology Facilities Council (STFC), United Kingdom;
National Science Foundation of the United States of America (NSF) and United States Department of Energy, Office of Nuclear Physics (DOE NP), United States of America.
In addition, individual groups or members have received support from:
Czech Science Foundation (grant no. 23-07499S), Czech Republic;
FORTE project, reg.\ no.\ CZ.02.01.01/00/22\_008/0004632, Czech Republic, co-funded by the European Union, Czech Republic;
European Research Council (grant no. 950692), European Union;
Deutsche Forschungs Gemeinschaft (DFG, German Research Foundation) ``Neutrinos and Dark Matter in Astro- and Particle Physics'' (grant no. SFB 1258), Germany;
ICSC - National Research Center for High Performance Computing, Big Data and Quantum Computing and FAIR - Future Artificial Intelligence Research, funded by the NextGenerationEU program (Italy).

\end{acknowledgement}

\bibliographystyle{utphys}   
\bibliography{bibliography.bib,alice_papers.bib}

@article{ALICE:2024gqr,
    author = "Acharya, Shreyasi and others",
    collaboration = "ALICE",
    title = "{Investigating charm quark energy loss in medium with the nuclear modification factor of D$^0$-tagged jets}",
    eprint = "2409.11939",
    archivePrefix = "arXiv",
    primaryClass = "nucl-ex",
    reportNumber = "CERN-EP-2024-240",
    month = "9",
    year = "2024"
}

@article{ALICE:2023qve,
    author = "Acharya, Shreyasi and others",
    collaboration = "ALICE",
    title = "{Observation of Medium-Induced Yield Enhancement and Acoplanarity Broadening of Low-$p_{\rm T}$ Jets from Measurements in pp and Central Pb-Pb Collisions at \sqrtsNN = 5.02 TeV}",
    eprint = "2308.16131",
    archivePrefix = "arXiv",
    primaryClass = "nucl-ex",
    reportNumber = "CERN-EP-2023-189",
    doi = "10.1103/PhysRevLett.133.022301",
    journal = "Phys. Rev. Lett.",
    volume = "133",
    number = "2",
    pages = "022301",
    year = "2024"
}

@article{ALICE:2023jye,
    author = "Acharya, Shreyasi and others",
    collaboration = "ALICE",
    title = "{Measurements of jet quenching using semi-inclusive hadron+jet distributions in pp and central Pb-Pb collisions at \sqrtsNN = 5.02 TeV}",
    eprint = "2308.16128",
    archivePrefix = "arXiv",
    primaryClass = "nucl-ex",
    reportNumber = "CERN-EP-2023-188",
    doi = "10.1103/PhysRevC.110.014906",
    journal = "Phys. Rev. C",
    volume = "110",
    number = "1",
    pages = "014906",
    year = "2024"
}

@article{Cortese:2019nnv,
    author = "Cortese, Pietro",
    collaboration = "ALICE",
    title = "{Performance of the ALICE Zero Degree Calorimeters and upgrade strategy}",
    doi = "10.1088/1742-6596/1162/1/012006",
    journal = "J. Phys. Conf. Ser.",
    volume = "1162",
    number = "1",
    pages = "012006",
    year = "2019"
}

@article{ALICE:2023kjg,
    author = "Acharya, Shreyasi and others",
    collaboration = "ALICE",
    title = "{Azimuthal correlations of heavy-flavor hadron decay electrons with charged particles in pp and p\textendash{}Pb collisions at $\sqrt{s_{\mathrm{{NN}}}}$ = 5.02 TeV}",
    eprint = "2303.00591",
    archivePrefix = "arXiv",
    primaryClass = "nucl-ex",
    reportNumber = "CERN-EP-2023-024",
    doi = "10.1140/epjc/s10052-023-11835-x",
    journal = "Eur. Phys. J. C",
    volume = "83",
    number = "8",
    pages = "741",
    year = "2023"
}

@article{ALICE:2018rtz,
    author = "Acharya, S. and others",
    collaboration = "ALICE",
    title = "{Energy dependence and fluctuations of anisotropic flow in Pb--Pb collisions at $ \sqrt{s_{\mathrm{NN}}}=5.02 $ and 2.76 TeV}",
    eprint = "1804.02944",
    archivePrefix = "arXiv",
    primaryClass = "nucl-ex",
    reportNumber = "CERN-EP-2018-057",
    doi = "10.1007/JHEP07(2018)103",
    journal = "JHEP",
    volume = "07",
    pages = "103",
    year = "2018"
}

@article{ALICE:2018vuu,
    author = "Acharya, S. and others",
    collaboration = "ALICE",
    title = "{Transverse momentum spectra and nuclear modification factors of charged particles in pp, p--Pb and Pb--Pb collisions at the LHC}",
    eprint = "1802.09145",
    archivePrefix = "arXiv",
    primaryClass = "nucl-ex",
    reportNumber = "CERN-EP-2018-025",
    doi = "10.1007/JHEP11(2018)013",
    journal = "JHEP",
    volume = "11",
    pages = "013",
    year = "2018"
}

@article{ALICE:2023xiu,
    author = "Acharya, Shreyasi and others",
    collaboration = "ALICE",
    title = "{Inclusive and multiplicity dependent production of electrons from heavy-flavour hadron decays in pp and p--Pb collisions}",
    eprint = "2303.13349",
    archivePrefix = "arXiv",
    primaryClass = "nucl-ex",
    reportNumber = "CERN-EP-2023-041",
    doi = "10.1007/JHEP08(2023)006",
    journal = "JHEP",
    volume = "08",
    pages = "006",
    year = "2023"
}

@article{ALICE:2023udb,
    author = "Acharya, Shreyasi and others",
    collaboration = "ALICE",
    title = "{ALICE upgrades during the LHC Long Shutdown 2}",
    eprint = "2302.01238",
    archivePrefix = "arXiv",
    primaryClass = "physics.ins-det",
    reportNumber = "CERN-EP-2023-009",
    doi = "10.1088/1748-0221/19/05/P05062",
    journal = "JINST",
    volume = "19",
    number = "05",
    pages = "P05062",
    year = "2024"
}

@article{ALICE:2022iba,
    author = "Acharya, Shreyasi and others",
    collaboration = "ALICE",
    title = "{Measurement of electrons from beauty-hadron decays in pp and Pb--Pb collisions at $\sqrt{s_{\mathrm{NN}}}=5.02$ TeV}",
    eprint = "2211.13985",
    archivePrefix = "arXiv",
    primaryClass = "nucl-ex",
    reportNumber = "CERN-EP-2022-238",
    doi = "10.1103/PhysRevC.108.034906",
    journal = "Phys. Rev. C",
    volume = "108",
    number = "3",
    pages = "034906",
    year = "2023"
}

@article{ALICE:2022wpn,
    author = "Acharya, Shreyasi and others",
    collaboration = "ALICE",
    title = "{The ALICE experiment: a journey through QCD}",
    eprint = "2211.04384",
    archivePrefix = "arXiv",
    primaryClass = "nucl-ex",
    reportNumber = "CERN-EP-2022-227",
    doi = "10.1140/epjc/s10052-024-12935-y",
    journal = "Eur. Phys. J. C",
    volume = "84",
    number = "8",
    pages = "813",
    year = "2024"
}

@article{ALICE:2022rau,
    author = "Acharya, Shreyasi and others",
    collaboration = "ALICE",
    title = "{Jet-like correlations with respect to K$^{0}_{\rm S}$ and $\Lambda$ ($\bar{\Lambda}$) in pp and Pb--Pb collisions at $\mathbf{\it\sqrt{s_\mathrm{NN}}}$ = 5.02 TeV}",
    eprint = "2211.01197",
    archivePrefix = "arXiv",
    primaryClass = "nucl-ex",
    reportNumber = "CERN-EP-2022-225",
    doi = "10.1140/epjc/s10052-023-11614-8",
    journal = "Eur. Phys. J. C",
    volume = "83",
    number = "6",
    pages = "497",
    year = "2023"
}

@article{ALICE:2022qhn,
    author = "Acharya, Shreyasi and others",
    collaboration = "ALICE",
    title = "{Performance of the ALICE Electromagnetic Calorimeter}",
    eprint = "2209.04216",
    archivePrefix = "arXiv",
    primaryClass = "physics.ins-det",
    reportNumber = "CERN-EP-2022-184",
    doi = "10.1088/1748-0221/18/08/P08007",
    journal = "JINST",
    volume = "18",
    number = "08",
    pages = "P08007",
    year = "2023"
}

@article{ALICE:2022tji,
    author = "Acharya, Shreyasi and others",
    collaboration = "ALICE",
    title = "{Measurement of beauty production via non-prompt D$^{0}$ mesons in Pb--Pb collisions at \sqrtsNN = 5.02 TeV}",
    eprint = "2202.00815",
    archivePrefix = "arXiv",
    primaryClass = "nucl-ex",
    reportNumber = "CERN-EP-2022-015",
    doi = "10.1007/JHEP12(2022)126",
    journal = "JHEP",
    volume = "12",
    pages = "126",
    year = "2022"
}

@article{ALICE:2021rxa,
    author = "Acharya, Shreyasi and others",
    collaboration = "ALICE",
    title = "{Prompt D$^{0}$, D$^{+}$, and D$^{*+}$ production in Pb\textendash{}Pb collisions at $ \sqrt{s_{\mathrm{NN}}} $ = 5.02 TeV}",
    eprint = "2110.09420",
    archivePrefix = "arXiv",
    primaryClass = "nucl-ex",
    reportNumber = "CERN-EP-2021-213",
    doi = "10.1007/JHEP01(2022)174",
    journal = "JHEP",
    volume = "01",
    pages = "174",
    year = "2022"
}

@article{ALICE:2019oyn,
    author = "Acharya, Shreyasi and others",
    collaboration = "ALICE",
    title = "{Azimuthal correlations of prompt D mesons with charged particles in pp and p\textendash{}Pb collisions at $\sqrt{s_{\mathrm{NN}}}$ = 5.02 TeV}",
    eprint = "1910.14403",
    archivePrefix = "arXiv",
    primaryClass = "nucl-ex",
    reportNumber = "CERN-EP-2019-239",
    doi = "10.1140/epjc/s10052-020-8118-0",
    journal = "Eur. Phys. J. C",
    volume = "80",
    number = "10",
    pages = "979",
    year = "2020"
}

@article{ALICE:2019bfx,
    author = "Acharya, Shreyasi and others",
    collaboration = "ALICE",
    title = "{Measurement of electrons from heavy-flavour hadron decays as a function of multiplicity in p--Pb collisions at $\sqrt{s_{\rm NN}}$ = 5.02 TeV}",
    eprint = "1910.14399",
    archivePrefix = "arXiv",
    primaryClass = "nucl-ex",
    reportNumber = "CERN-EP-2019-250",
    doi = "10.1007/JHEP02(2020)077",
    journal = "JHEP",
    volume = "02",
    pages = "077",
    year = "2020"
}

@article{ALICE:2019nuy,
    author = "Acharya, Shreyasi and others",
    collaboration = "ALICE",
    title = "{Measurement of electrons from semileptonic heavy-flavour hadron decays at midrapidity in pp and Pb--Pb collisions at $\sqrt{s_{\rm{NN}}}$ = 5.02 TeV}",
    eprint = "1910.09110",
    archivePrefix = "arXiv",
    primaryClass = "nucl-ex",
    reportNumber = "CERN-EP-2019-205",
    doi = "10.1016/j.physletb.2020.135377",
    journal = "Phys. Lett. B",
    volume = "804",
    pages = "135377",
    year = "2020"
}

@article{ALICE:2018yau,
    author = "Acharya, Shreyasi and others",
    collaboration = "ALICE",
    title = "{Measurements of low-p$_{T}$ electrons from semileptonic heavy-flavour hadron decays at mid-rapidity in pp and Pb--Pb collisions at $ \sqrt{s_{\mathrm{NN}}}=2.76 $ TeV}",
    eprint = "1805.04379",
    archivePrefix = "arXiv",
    primaryClass = "nucl-ex",
    reportNumber = "CERN-EP-2018-095",
    doi = "10.1007/JHEP10(2018)061",
    journal = "JHEP",
    volume = "10",
    pages = "061",
    year = "2018"
}

@article{ALICE:2018lyv,
    author = "Acharya, S. and others",
    collaboration = "ALICE",
    title = "{Measurement of D$^{0}$, D$^{+}$, D$^{*+}$ and D$_{\rm{s}}^{+}$ production in Pb--Pb collisions at $ \sqrt{{\mathrm{s}}_{\mathrm{NN}}}=5.02 $ TeV}",
    eprint = "1804.09083",
    archivePrefix = "arXiv",
    primaryClass = "nucl-ex",
    reportNumber = "CERN-EP-2018-066",
    doi = "10.1007/JHEP10(2018)174",
    journal = "JHEP",
    volume = "10",
    pages = "174",
    year = "2018"
}

@article{ALICE:2016rzo,
    author = "Adam, Jaroslav and others",
    collaboration = "ALICE",
    title = "{W and Z boson production in p--Pb collisions at $\sqrt{s_{\rm NN}}$ = 5.02 TeV}",
    eprint = "1611.03002",
    archivePrefix = "arXiv",
    primaryClass = "nucl-ex",
    reportNumber = "CERN-EP-2016-278",
    doi = "10.1007/JHEP02(2017)077",
    journal = "JHEP",
    volume = "02",
    pages = "077",
    year = "2017"
}

@article{ALICE:2016mpw,
    author = "Adam, Jaroslav and others",
    collaboration = "ALICE",
    title = "{Measurement of the production of high-$p_{\rm T}$ electrons from heavy-flavour hadron decays in Pb--Pb collisions at $\mathbf{\sqrt{\it s_{\rm{NN}}}}$ = 2.76 TeV}",
    eprint = "1609.07104",
    archivePrefix = "arXiv",
    primaryClass = "nucl-ex",
    reportNumber = "CERN-EP-2016-235",
    doi = "10.1016/j.physletb.2017.05.060",
    journal = "Phys. Lett. B",
    volume = "771",
    pages = "467--481",
    year = "2017"
}

@article{ALICE:2016gso,
    author = "Adam, Jaroslav and others",
    collaboration = "ALICE",
    title = "{Jet-like correlations with neutral pion triggers in pp and central Pb\textendash{}Pb collisions at $\sqrt{s_{\mathrm{NN}}}=2.76$ TeV}",
    eprint = "1608.07201",
    archivePrefix = "arXiv",
    primaryClass = "nucl-ex",
    reportNumber = "CERN-EP-2016-195",
    doi = "10.1016/j.physletb.2016.10.048",
    journal = "Phys. Lett. B",
    volume = "763",
    pages = "238--250",
    year = "2016"
}

@article{ALICE:2016clc,
    author = "Adam, Jaroslav and others",
    collaboration = "ALICE",
    title = "{Measurement of azimuthal correlations of D mesons and charged particles in pp collisions at $\sqrt{s}=7$ TeV and p--Pb collisions at $\sqrt{s_{\rm NN}}=5.02$ TeV}",
    eprint = "1605.06963",
    archivePrefix = "arXiv",
    primaryClass = "nucl-ex",
    reportNumber = "CERN-EP-2016-129",
    doi = "10.1140/epjc/s10052-017-4779-8",
    journal = "Eur. Phys. J. C",
    volume = "77",
    number = "4",
    pages = "245",
    year = "2017"
}

@article{ALICE:2015zhm,
    author = "Adam, Jaroslav and others",
    collaboration = "ALICE",
    title = "{Measurement of electrons from heavy-flavour hadron decays in p--Pb collisions at $\sqrt{s_{\rm NN}} =$ 5.02 TeV}",
    eprint = "1509.07491",
    archivePrefix = "arXiv",
    primaryClass = "nucl-ex",
    reportNumber = "CERN-PH-EP-2015-262",
    doi = "10.1016/j.physletb.2015.12.067",
    journal = "Phys. Lett. B",
    volume = "754",
    pages = "81--93",
    year = "2016"
}

@article{ALICE:2014aev,
    author = "Abelev, Betty Bezverkhny and others",
    collaboration = "ALICE",
    title = "{Beauty production in pp collisions at $\sqrt{s}$ = 2.76 TeV measured via semi-electronic decays}",
    eprint = "1405.4144",
    archivePrefix = "arXiv",
    primaryClass = "nucl-ex",
    reportNumber = "CERN-PH-EP-2014-089",
    doi = "10.1016/j.physletb.2014.09.026",
    journal = "Phys. Lett. B",
    volume = "738",
    pages = "97--108",
    year = "2014"
}

@article{ALICE:2014sbx,
    author = "Abelev, Betty Bezverkhny and others",
    collaboration = "ALICE",
    title = "{Performance of the ALICE Experiment at the CERN LHC}",
    eprint = "1402.4476",
    archivePrefix = "arXiv",
    primaryClass = "nucl-ex",
    reportNumber = "CERN-PH-EP-2014-031",
    doi = "10.1142/S0217751X14300440",
    journal = "Int. J. Mod. Phys. A",
    volume = "29",
    pages = "1430044",
    year = "2014"
}

@article{ALICE:2013snk,
    author = "Abelev, Betty Bezverkhny and others",
    collaboration = "ALICE",
    title = "{Long-range angular correlations of $\rm \pi$, K and p in p--Pb collisions at $\sqrt{s_{\rm NN}}$ = 5.02 TeV}",
    eprint = "1307.3237",
    archivePrefix = "arXiv",
    primaryClass = "nucl-ex",
    reportNumber = "CERN-PH-EP-2013-115",
    doi = "10.1016/j.physletb.2013.08.024",
    journal = "Phys. Lett. B",
    volume = "726",
    pages = "164--177",
    year = "2013"
}

@article{ALICE:2013axi,
    author = "Abbas, E. and others",
    collaboration = "ALICE",
    title = "{Performance of the ALICE VZERO system}",
    eprint = "1306.3130",
    archivePrefix = "arXiv",
    primaryClass = "nucl-ex",
    reportNumber = "CERN-PH-EP-2013-082",
    doi = "10.1088/1748-0221/8/10/P10016",
    journal = "JINST",
    volume = "8",
    pages = "P10016",
    year = "2013"
}

@article{ALICE:2012mzy,
    author = "Abelev, Betty and others",
    collaboration = "ALICE",
    title = "{Measurement of electrons from semileptonic heavy-flavour hadron decays in pp collisions at $\sqrt{s} = 7$ TeV}",
    eprint = "1205.5423",
    archivePrefix = "arXiv",
    primaryClass = "hep-ex",
    reportNumber = "CERN-PH-EP-2012-131",
    doi = "10.1103/PhysRevD.86.112007",
    journal = "Phys. Rev. D",
    volume = "86",
    pages = "112007",
    year = "2012"
}

@article{ALICE:2012ab,
    author = "Abelev, Betty and others",
    collaboration = "ALICE",
    title = "{Suppression of high transverse momentum D mesons in central Pb--Pb collisions at $\sqrt{s_{\mathrm{NN}}}=2.76$ TeV}",
    eprint = "1203.2160",
    archivePrefix = "arXiv",
    primaryClass = "nucl-ex",
    reportNumber = "CERN-PH-EP-2012-069",
    doi = "10.1007/JHEP09(2012)112",
    journal = "JHEP",
    volume = "09",
    pages = "112",
    year = "2012"
}

@article{ALICE:2011gpa,
    author = "Aamodt, K. and others",
    collaboration = "ALICE",
    title = "{Particle-yield modification in jet-like azimuthal di-hadron correlations in Pb--Pb collisions at $\sqrt{s_{\mathrm{NN}}} = 2.76$ TeV}",
    eprint = "1110.0121",
    archivePrefix = "arXiv",
    primaryClass = "nucl-ex",
    reportNumber = "CERN-PH-EP-2011-161",
    doi = "10.1103/PhysRevLett.108.092301",
    journal = "Phys. Rev. Lett.",
    volume = "108",
    pages = "092301",
    year = "2012"
}

@article{ALICE:2011svq,
    author = "Aamodt, K. and others",
    collaboration = "ALICE",
    title = "{Harmonic decomposition of two-particle angular correlations in Pb--Pb collisions at $\sqrt{s_{\mathrm{NN}}}=$ 2.76 TeV}",
    eprint = "1109.2501",
    archivePrefix = "arXiv",
    primaryClass = "nucl-ex",
    reportNumber = "CERN-PH-EP-2011-152",
    doi = "10.1016/j.physletb.2012.01.060",
    journal = "Phys. Lett. B",
    volume = "708",
    pages = "249--264",
    year = "2012"
}

@article{ALICE:2010tia,
    author = "Aamodt, K and others",
    collaboration = "ALICE",
    title = "{Alignment of the ALICE Inner Tracking System with cosmic-ray tracks}",
    eprint = "1001.0502",
    archivePrefix = "arXiv",
    primaryClass = "physics.ins-det",
    doi = "10.1088/1748-0221/5/03/P03003",
    journal = "JINST",
    volume = "5",
    pages = "P03003",
    year = "2010"
}

@article{CMS:2019jis,
    author = "Sirunyan, Albert M and others",
    collaboration = "CMS",
    title = "{Studies of charm quark diffusion inside jets using PbPb and pp collisions at $\sqrt{s_\mathrm{NN}} =$ 5.02 TeV}",
    eprint = "1911.01461",
    archivePrefix = "arXiv",
    primaryClass = "hep-ex",
    reportNumber = "CMS-HIN-18-007, CERN-EP-2019-247",
    doi = "10.1103/PhysRevLett.125.102001",
    journal = "Phys. Rev. Lett.",
    volume = "125",
    number = "10",
    pages = "102001",
    year = "2020"
}

@article{CMS:2024krd,
    author = "Hayrapetyan, Aram and others",
    collaboration = "CMS",
    title = "{Overview of high-density QCD studies with the CMS experiment at the LHC}",
    eprint = "2405.10785",
    archivePrefix = "arXiv",
    primaryClass = "nucl-ex",
    reportNumber = "CMS-HIN-23-011, CERN-EP-2024-057",
    doi = "10.1016/j.physrep.2024.11.007",
    journal = "Phys. Rept.",
    volume = "1115",
    pages = "219--367",
    year = "2025"
}

@article{Cacciari:2012ny,
    author = "Cacciari, Matteo and Frixione, Stefano and Houdeau, Nicolas and Mangano, Michelangelo L. and Nason, Paolo and Ridolfi, Giovanni",
    title = "{Theoretical predictions for charm and bottom production at the LHC}",
    eprint = "1205.6344",
    archivePrefix = "arXiv",
    primaryClass = "hep-ph",
    reportNumber = "CERN-PH-TH-2011-227",
    doi = "10.1007/JHEP10(2012)137",
    journal = "JHEP",
    volume = "10",
    pages = "137",
    year = "2012"
}

@article{Yang:2025dqu,
    author = "Yang, Zhong and Wang, Xin-Nian",
    title = "{Rapidity asymmetry of jet-hadron correlation as a robust signal of diffusion wake induced by di-jets in high-energy heavy-ion collisions}",
    eprint = "2501.03419",
    archivePrefix = "arXiv",
    primaryClass = "hep-ph",
    month = "1",
    year = "2025"
}

@article{Yang:2022nei,
    author = "Yang, Zhong and Luo, Tan and Chen, Wei and Pang, Long-Gang and Wang, Xin-Nian",
    title = "{3D Structure of Jet-Induced Diffusion Wake in an Expanding Quark-Gluon Plasma}",
    eprint = "2203.03683",
    archivePrefix = "arXiv",
    primaryClass = "hep-ph",
    doi = "10.1103/PhysRevLett.130.052301",
    journal = "Phys. Rev. Lett.",
    volume = "130",
    number = "5",
    pages = "052301",
    year = "2023"
}

@article{Chen:2023dzi,
    author = "Chen, W. and Yang, Z. and Pang, L. and He, Y. and Luo, T. and Wang, X. N.",
    title = "{The Signals of Jet-induced Diffusion Wake on $Z/\gamma $-hadron Correlations in High-Energy Heavy-Ion Collisions}",
    doi = "10.5506/APhysPolBSupp.16.1-A53",
    journal = "Acta Phys. Polon. Supp.",
    volume = "16",
    number = "1",
    pages = "1--A53",
    year = "2023"
}

@article{Vitev:2002pf,
    author = "Vitev, Ivan and Gyulassy, Miklos",
    title = "{High $p_\mathrm{T}$ tomography of $d$ + Au and Au+Au at SPS, RHIC, and LHC}",
    eprint = "hep-ph/0209161",
    archivePrefix = "arXiv",
    doi = "10.1103/PhysRevLett.89.252301",
    journal = "Phys. Rev. Lett.",
    volume = "89",
    pages = "252301",
    year = "2002"
}

@article{Kopeliovich:2002yh,
    author = "Kopeliovich, B. Z. and Nemchik, J. and Schafer, A. and Tarasov, A. V.",
    title = "{Cronin effect in hadron production off nuclei}",
    eprint = "hep-ph/0201010",
    archivePrefix = "arXiv",
    doi = "10.1103/PhysRevLett.88.232303",
    journal = "Phys. Rev. Lett.",
    volume = "88",
    pages = "232303",
    year = "2002"
}

@article{Wang:1998ww,
    author = "Wang, Xin-Nian",
    title = "{Systematic study of high $p_{T}$ hadron spectra in $p p$, $p$ A and A A collisions from SPS to RHIC energies}",
    eprint = "nucl-th/9812021",
    archivePrefix = "arXiv",
    reportNumber = "LBNL-42545, LBL-42545",
    doi = "10.1103/PhysRevC.61.064910",
    journal = "Phys. Rev. C",
    volume = "61",
    pages = "064910",
    year = "2000"
}

@article{Ma:2010dv,
    author = "Ma, Guo-Liang and Wang, Xin-Nian",
    title = "{Jets, Mach cone, hot spots, ridges, harmonic flow, dihadron and $\gamma$-hadron correlation in high-energy heavy-ion collisions}",
    eprint = "1011.5249",
    archivePrefix = "arXiv",
    primaryClass = "nucl-th",
    doi = "10.1103/PhysRevLett.106.162301",
    journal = "Phys. Rev. Lett.",
    volume = "106",
    pages = "162301",
    year = "2011"
}

@article{Vitev:2005yg,
    author = "Vitev, Ivan",
    title = "{Large angle hadron correlations from medium-induced gluon radiation}",
    eprint = "hep-ph/0501255",
    archivePrefix = "arXiv",
    doi = "10.1016/j.physletb.2005.09.082",
    journal = "Phys. Lett. B",
    volume = "630",
    pages = "78--84",
    year = "2005"
}

@article{Li:2024pfi,
    author = "Li, Yao and Chen, Shi-Yong and Kong, Wei-Xi and Wang, Sa and Zhang, Ben-Wei",
    title = "{Medium Modifications of Heavy-Flavor Jet Angularities in High-Energy Nuclear Collisions}",
    eprint = "2409.12742",
    archivePrefix = "arXiv",
    primaryClass = "hep-ph",
    doi = "10.1088/0256-307X/42/1/011201",
    journal = "Chin. Phys. Lett.",
    volume = "42",
    number = "1",
    pages = "011201",
    year = "2025"
}

@article{Cao:2022odi,
    author = "Cao, Shanshan and Qin, Guang-You",
    title = "{Medium Response and Jet\textendash{}Hadron Correlations in Relativistic Heavy-Ion Collisions}",
    eprint = "2211.16821",
    archivePrefix = "arXiv",
    primaryClass = "nucl-th",
    doi = "10.1146/annurev-nucl-112822-031317",
    journal = "Ann. Rev. Nucl. Part. Sci.",
    volume = "73",
    pages = "205--229",
    year = "2023"
}

@article{PHENIX:2008osq,
    author = "Adare, A. and others",
    collaboration = "PHENIX",
    title = "{Dihadron azimuthal correlations in Au$+$Au collisions at \sqrtsNN = 200 GeV}",
    eprint = "0801.4545",
    archivePrefix = "arXiv",
    primaryClass = "nucl-ex",
    doi = "10.1103/PhysRevC.78.014901",
    journal = "Phys. Rev. C",
    volume = "78",
    pages = "014901",
    year = "2008"
}

@article{CMS:2014jjt,
    author = "Chatrchyan, Serguei and others",
    collaboration = "CMS",
    title = "{Measurement of Jet Fragmentation in PbPb and pp Collisions at $\sqrt{s_\mathrm{NN}}= 2.76$ TeV}",
    eprint = "1406.0932",
    archivePrefix = "arXiv",
    primaryClass = "nucl-ex",
    reportNumber = "CMS-HIN-12-013, CERN-PH-EP-2014-100",
    doi = "10.1103/PhysRevC.90.024908",
    journal = "Phys. Rev. C",
    volume = "90",
    number = "2",
    pages = "024908",
    year = "2014"
}

@article{ParticleDataGroup:2024cfk,
    author = "Navas, S. and others",
    collaboration = "Particle Data Group",
    title = "{Review of particle physics}",
    doi = "10.1103/PhysRevD.110.030001",
    journal = "Phys. Rev. D",
    volume = "110",
    number = "3",
    pages = "030001",
    year = "2024"
}

@article{fishersphere,
author = {Fisher, Ronald Aylmer },
title = {Dispersion on a sphere},
journal = {Proceedings of the Royal Society of London. Series A. Mathematical and Physical Sciences},
volume = {217},
number = {1130},
pages = {295-305},
year = {1953},
doi = {10.1098/rspa.1953.0064},
URL = {https://royalsocietypublishing.org/doi/abs/10.1098/rspa.1953.0064},
}

@article{AWES1992130,
title = {A simple method of shower localization and identification in laterally segmented calorimeters},
journal = {Nuclear Instruments and Methods in Physics Research Section A: Accelerators, Spectrometers, Detectors and Associated Equipment},
volume = {311},
number = {1},
pages = {130-138},
year = {1992},
issn = {0168-9002},
doi = {https://doi.org/10.1016/0168-9002(92)90858-2},
author = {T.C. Awes and F.E. Obenshain and F. Plasil and S. Saini and S.P. Sorensen and G.R. Young}
}

@article{ATLAS:2020yxw,
    author = "Aad, Georges and others",
    collaboration = "ATLAS",
    title = "{Measurement of azimuthal anisotropy of muons from charm and bottom hadrons in Pb+Pb collisions at $\sqrt{s_\mathrm{NN}}=5.02$ TeV with the ATLAS detector}",
    eprint = "2003.03565",
    archivePrefix = "arXiv",
    primaryClass = "nucl-ex",
    reportNumber = "CERN-EP-2019-274",
    doi = "10.1016/j.physletb.2020.135595",
    journal = "Phys. Lett. B",
    volume = "807",
    pages = "135595",
    year = "2020"
}

@article{ATLAS:2021xtw,
    author = "Aad, Georges and others",
    collaboration = "ATLAS",
    title = "{Measurement of the nuclear modification factor for muons from charm and bottom hadrons in Pb+Pb collisions at 5.02 TeV with the ATLAS detector}",
    eprint = "2109.00411",
    archivePrefix = "arXiv",
    primaryClass = "nucl-ex",
    reportNumber = "CERN-EP-2021-153",
    doi = "10.1016/j.physletb.2022.137077",
    journal = "Phys. Lett. B",
    volume = "829",
    pages = "137077",
    year = "2022"
}

@article{ALICE-PUBLIC-2017-005,
      author        = "S. Acharya and others",
      title         = "{The ALICE definition of primary particles}",
      collaboration = "ALICE",
      number        = "ALICE-PUBLIC-2017-005",
      month         = "Jun",
      year          = "2017",
      report    = "\href{https://cds.cern.ch/record/2270008}{ALICE-PUBLIC-2017-005}"
}

@article{Bethe:1930ku,
    author = "Bethe, H.",
    title = "{Theory of the Passage of Fast Corpuscular Rays Through Matter}",
    doi = "10.1002/andp.19303970303",
    journal = "Annalen Phys.",
    volume = "5",
    pages = "325--400",
    year = "1930"
}

@article{Alme:2010ke,
    author = "Alme, J. and others",
    title = "{The ALICE TPC, a large 3-dimensional tracking device with fast readout for ultra-high multiplicity events}",
    eprint = "1001.1950",
    archivePrefix = "arXiv",
    primaryClass = "physics.ins-det",
    doi = "10.1016/j.nima.2010.04.042",
    journal = "Nucl. Instrum. Meth. A",
    volume = "622",
    pages = "316--367",
    year = "2010"
}

@article{Cortese:1121574,
      author        = "Cortese, P. and others",
      title         = "{ALICE Electromagnetic Calorimeter Technical Design Report}",
      collaboration = "ALICE",
      number        = "CERN-LHCC-2008-014, ALICE-TDR-14",
      month         = "Aug",
      year          = "2008",
      report  = "\href{http://cds.cern.ch/record/1121574}{CERN-LHCC-2008-014}"
}

@article{Allen:2010stl,
      author         = "Allen, J. and others",
      title          = "{ALICE DCal: An Addendum to the EMCal Technical Design Report Di-Jet and Hadron-Jet correlation measurements in ALICE}",
      collaboration  = "ALICE",
      number        = "CERN-LHCC-2010-011, ALICE-TDR-14-add-1",
      year           = "2010",
      report   = "\href{http://cds.cern.ch/record/1272952}{CERN-LHCC-2010-011, ALICE-TDR-14-add-1}"
}

@article{Aamodt:2008zz,
    author = "Aamodt, K. and others",
    collaboration = "ALICE",
    title = "{The ALICE experiment at the CERN LHC}",
    doi = "10.1088/1748-0221/3/08/S08002",
    journal = "JINST",
    volume = "3",
    pages = "S08002",
    year = "2008"
}

@article{ALICE:2018tvk,
    author        = "S. Acharya and others",
    title = "{Centrality determination in heavy ion collisions}",
    collaboration = "ALICE",
    number = "ALICE-PUBLIC-2018-011",
    month = "Aug",
    year = "2018",
    report = "\href{https://cds.cern.ch/record/2636623}{ALICE-PUBLIC-2018-011}"
}

@article{Sjostrand:2006za,
    author = "Sjostrand, Torbjorn and Mrenna, Stephen and Skands, Peter Z.",
    title = "{PYTHIA 6.4 Physics and Manual}",
    eprint = "hep-ph/0603175",
    archivePrefix = "arXiv",
    reportNumber = "FERMILAB-PUB-06-052-CD-T, LU-TP-06-13",
    doi = "10.1088/1126-6708/2006/05/026",
    journal = "JHEP",
    volume = "05",
    pages = "026",
    year = "2006"
}

@article{Wang:1991hta,
    author = "Wang, Xin-Nian and Gyulassy, Miklos",
    title = "{HIJING: A Monte Carlo model for multiple jet production in pp, pA and AA collisions}",
    reportNumber = "LBL-31036",
    doi = "10.1103/PhysRevD.44.3501",
    journal = "Phys. Rev. D",
    volume = "44",
    pages = "3501--3516",
    year = "1991"
}

@book{Brun:1073159,
      author        = "Brun, René and Bruyant, F and Carminati, Federico and
                       Giani, Simone and Maire, M and McPherson, A and Patrick, G
                       and Urban, L",
      title         = "{GEANT: detector description and simulation tool; March
                       1994}",
      publisher     = "CERN",
      address       = "Geneva",
      series        = "CERN Program Library",
      year          = "1993",
      url           = "http://cds.cern.ch/record/1073159",
      note          = "Long Writeup W5013",
}

@article{ALEPH:1994dex,
  author       = {Assmann, R.W. and others},
  title        = {{Calibration of the ALEPH dE/dx}},
  collaboration  = {ALEPH},
  year         = {1994},
  report       = "\href{https://cds.cern.ch/record/806004}{ALEPH 94-116, PHYSIC 94-089}",
  number       = {ALEPH 94-116, PHYSIC 94-089},
  note         = {CERN Document Server}
}

@article{Borsanyi:2010cj,
    author = "Borsanyi, Szabolcs and Endrodi, Gergely and Fodor, Zoltan and Jakovac, Antal and Katz, Sandor D. and Krieg, Stefan and Ratti, Claudia and Szabo, Kalman K.",
    title = "{The QCD equation of state with dynamical quarks}",
    eprint = "1007.2580",
    archivePrefix = "arXiv",
    primaryClass = "hep-lat",
    reportNumber = "WUB-10-12",
    doi = "10.1007/JHEP11(2010)077",
    journal = "JHEP",
    volume = "11",
    pages = "077",
    year = "2010"
}

@article{ALEPH:1994ayc,
    author = "Buskulic, D. and others",
    collaboration = "ALEPH",
    title = "{Performance of the ALEPH detector at LEP}",
    reportNumber = "CERN-PPE-94-170, FSU-SCRI-95-70",
    doi = "10.1016/0168-9002(95)00138-7",
    journal = "Nucl. Instrum. Meth. A",
    volume = "360",
    pages = "481--506",
    year = "1995"
}

@article{ATLAS:2018bvp,
    author = "Aaboud, Morad and others",
    collaboration = "ATLAS",
    title = "{Measurement of jet fragmentation in Pb+Pb and $pp$ collisions at \sqrtsNN = 5.02 TeV with the ATLAS detector}",
    eprint = "1805.05424",
    archivePrefix = "arXiv",
    primaryClass = "nucl-ex",
    reportNumber = "CERN-EP-2018-096",
    doi = "10.1103/PhysRevC.98.024908",
    journal = "Phys. Rev. C",
    volume = "98",
    number = "2",
    pages = "024908",
    year = "2018"
}

@article{Ipek:2018lhm,
    author = "Ipek, Seyda and Tait, Tim M. P.",
    title = "{Early Cosmological Period of QCD Confinement}",
    eprint = "1811.00559",
    archivePrefix = "arXiv",
    primaryClass = "hep-ph",
    reportNumber = "UCI-HEP-TR-2018-15",
    doi = "10.1103/PhysRevLett.122.112001",
    journal = "Phys. Rev. Lett.",
    volume = "122",
    number = "11",
    pages = "112001",
    year = "2019"
}

@article{Bazavov:2011nk,
    author = "Bazavov, A. and others",
    title = "{The chiral and deconfinement aspects of the QCD transition}",
    eprint = "1111.1710",
    archivePrefix = "arXiv",
    primaryClass = "hep-lat",
    doi = "10.1103/PhysRevD.85.054503",
    journal = "Phys. Rev. D",
    volume = "85",
    pages = "054503",
    year = "2012"
}

@article{Bazavov:2018mes,
    author = "Bazavov, A. and others",
    collaboration = "HotQCD",
    title = "{Chiral crossover in QCD at zero and non-zero chemical potentials}",
    eprint = "1812.08235",
    archivePrefix = "arXiv",
    primaryClass = "hep-lat",
    doi = "10.1016/j.physletb.2019.05.013",
    journal = "Phys. Lett. B",
    volume = "795",
    pages = "15--21",
    year = "2019"
}

@article{Bazavov:2009zn,
    author = "Bazavov, A. and others",
    title = "{Equation of state and QCD transition at finite temperature}",
    eprint = "0903.4379",
    archivePrefix = "arXiv",
    primaryClass = "hep-lat",
    reportNumber = "CU-TP-1189",
    doi = "10.1103/PhysRevD.80.014504",
    journal = "Phys. Rev. D",
    volume = "80",
    pages = "014504",
    year = "2009"
}

@article{Muller:2012zq,
    author = "Muller, Berndt and Schukraft, Jurgen and Wyslouch, Boleslaw",
    title = "{First Results from Pb+Pb collisions at the LHC}",
    eprint = "1202.3233",
    archivePrefix = "arXiv",
    primaryClass = "hep-ex",
    reportNumber = "CERN-OPEN-2012-005",
    doi = "10.1146/annurev-nucl-102711-094910",
    journal = "Ann. Rev. Nucl. Part. Sci.",
    volume = "62",
    pages = "361--386",
    year = "2012"
}

@article{Arsene:2004fa,
    author = "Arsene, I. and others",
    collaboration = "BRAHMS",
    title = "{Quark gluon plasma and color glass condensate at RHIC? The Perspective from the BRAHMS experiment}",
    eprint = "nucl-ex/0410020",
    archivePrefix = "arXiv",
    doi = "10.1016/j.nuclphysa.2005.02.130",
    journal = "Nucl. Phys. A",
    volume = "757",
    pages = "1--27",
    year = "2005"
}

@article{Back:2004je,
    author = "Back, B. B. and others",
    collaboration = "PHOBOS",
    title = "{The PHOBOS perspective on discoveries at RHIC}",
    eprint = "nucl-ex/0410022",
    archivePrefix = "arXiv",
    doi = "10.1016/j.nuclphysa.2005.03.084",
    journal = "Nucl. Phys. A",
    volume = "757",
    pages = "28--101",
    year = "2005"
}

@article{Adams:2005dq,
    author = "Adams, John and others",
    collaboration = "STAR",
    title = "{Experimental and theoretical challenges in the search for the quark gluon plasma: The STAR Collaboration's critical assessment of the evidence from RHIC collisions}",
    eprint = "nucl-ex/0501009",
    archivePrefix = "arXiv",
    doi = "10.1016/j.nuclphysa.2005.03.085",
    journal = "Nucl. Phys. A",
    volume = "757",
    pages = "102--183",
    year = "2005"
}

@article{Adcox:2004mh,
    author = "Adcox, K. and others",
    collaboration = "PHENIX",
    title = "{Formation of dense partonic matter in relativistic nucleus-nucleus collisions at RHIC: Experimental evaluation by the PHENIX collaboration}",
    eprint = "nucl-ex/0410003",
    archivePrefix = "arXiv",
    doi = "10.1016/j.nuclphysa.2005.03.086",
    journal = "Nucl. Phys. A",
    volume = "757",
    pages = "184--283",
    year = "2005"
}

@article{Baier:2000mf,
    author = "Baier, R. and Schiff, D. and Zakharov, B. G.",
    title = "{Energy loss in perturbative QCD}",
    eprint = "hep-ph/0002198",
    archivePrefix = "arXiv",
    reportNumber = "BI-TP-2000-08, LPT-ORSAY-00-22",
    doi = "10.1146/annurev.nucl.50.1.37",
    journal = "Ann. Rev. Nucl. Part. Sci.",
    volume = "50",
    pages = "37--69",
    year = "2000"
}

@article{Dokshitzer:2001zm,
    author = "Dokshitzer, Yuri L. and Kharzeev, D. E.",
    title = "{Heavy quark colorimetry of QCD matter}",
    eprint = "hep-ph/0106202",
    archivePrefix = "arXiv",
    reportNumber = "LPT-ORSAY-01-58, BNL-NT-01-9",
    doi = "10.1016/S0370-2693(01)01130-3",
    journal = "Phys. Lett. B",
    volume = "519",
    pages = "199--206",
    year = "2001"
}

@article{Armesto:2003jh,
    author = "Armesto, Nestor and Salgado, Carlos A. and Wiedemann, Urs Achim",
    title = "{Medium induced gluon radiation off massive quarks fills the dead cone}",
    eprint = "hep-ph/0312106",
    archivePrefix = "arXiv",
    reportNumber = "CERN-TH-2003-301",
    doi = "10.1103/PhysRevD.69.114003",
    journal = "Phys. Rev. D",
    volume = "69",
    pages = "114003",
    year = "2004"
}

@article{Wicks:2007am,
    author = "Wicks, Simon and Horowitz, William and Djordjevic, Magdalena and Gyulassy, Miklos",
    editor = "Jacobs, P. and Wang, Xin-Nian",
    title = "{Heavy quark jet quenching with collisional plus radiative energy loss and path length fluctuations}",
    eprint = "nucl-th/0701063",
    archivePrefix = "arXiv",
    doi = "10.1016/j.nuclphysa.2006.11.102",
    journal = "Nucl. Phys. A",
    volume = "783",
    pages = "493--496",
    year = "2007"
}

@article{Zhang:2003wk,
    author = "Zhang, Ben-Wei and Wang, Enke and Wang, Xin-Nian",
    title = "{Heavy quark energy loss in nuclear medium}",
    eprint = "nucl-th/0309040",
    archivePrefix = "arXiv",
    reportNumber = "LBNL-53791",
    doi = "10.1103/PhysRevLett.93.072301",
    journal = "Phys. Rev. Lett.",
    volume = "93",
    pages = "072301",
    year = "2004"
}

@article{Adil:2006ra,
    author = "Adil, Azfar and Vitev, Ivan",
    title = "{Collisional dissociation of heavy mesons in dense QCD matter}",
    eprint = "hep-ph/0611109",
    archivePrefix = "arXiv",
    doi = "10.1016/j.physletb.2007.03.050",
    journal = "Phys. Lett. B",
    volume = "649",
    pages = "139--146",
    year = "2007"
}

@article{Cacciari:2008gp,
    author = "Cacciari, Matteo and Salam, Gavin P. and Soyez, Gregory",
    title = "{The anti-$k_t$ jet clustering algorithm}",
    eprint = "0802.1189",
    archivePrefix = "arXiv",
    primaryClass = "hep-ph",
    reportNumber = "LPTHE-07-03",
    doi = "10.1088/1126-6708/2008/04/063",
    journal = "JHEP",
    volume = "04",
    pages = "063",
    year = "2008"
}

@article{Cacciari:2011ma,
    author = "Cacciari, Matteo and Salam, Gavin P. and Soyez, Gregory",
    title = "{FastJet User Manual}",
    eprint = "1111.6097",
    archivePrefix = "arXiv",
    primaryClass = "hep-ph",
    reportNumber = "CERN-PH-TH-2011-297",
    doi = "10.1140/epjc/s10052-012-1896-2",
    journal = "Eur. Phys. J. C",
    volume = "72",
    pages = "1896",
    year = "2012"
}

@article{Casalderrey-Solana:2014bpa,
    author = "Casalderrey-Solana, Jorge and Gulhan, Doga Can and Milhano, Jos\'e Guilherme and Pablos, Daniel and Rajagopal, Krishna",
    title = "{A Hybrid Strong/Weak Coupling Approach to Jet Quenching}",
    eprint = "1405.3864",
    archivePrefix = "arXiv",
    primaryClass = "hep-ph",
    reportNumber = "MIT-CTP-4550, CERN-PH-TH-2014-089, ICCUB-14-051",
    doi = "10.1007/JHEP09(2015)175",
    journal = "JHEP",
    volume = "10",
    pages = "019",
    year = "2014",
    note = "[Erratum: JHEP 09, 175 (2015)]"
}

@article{Blaizot:2014ula,
    author = "Blaizot, Jean-Paul and Mehtar-Tani, Yacine and Torres, Marcus A. C.",
    title = "{Angular structure of the in-medium QCD cascade}",
    eprint = "1407.0326",
    archivePrefix = "arXiv",
    primaryClass = "hep-ph",
    doi = "10.1103/PhysRevLett.114.222002",
    journal = "Phys. Rev. Lett.",
    volume = "114",
    number = "22",
    pages = "222002",
    year = "2015"
}

@article{ATLAS:2022agz,
    author = "Aad, Georges and others",
    collaboration = "ATLAS",
    title = "{Measurement of the nuclear modification factor of $b$-jets in 5.02~TeV Pb+Pb collisions with the ATLAS detector}",
    eprint = "2204.13530",
    archivePrefix = "arXiv",
    primaryClass = "nucl-ex",
    reportNumber = "CERN-EP-2022-043",
    doi = "10.1140/epjc/s10052-023-11427-9",
    journal = "Eur. Phys. J. C",
    volume = "83",
    number = "5",
    pages = "438",
    year = "2023"
}

@article{CMS:2017qjw,
    author = "Sirunyan, Albert M and others",
    collaboration = "CMS",
    title = "{Nuclear modification factor of D$^0$ mesons in PbPb collisions at  $\sqrt{s_\mathrm{NN}} = 5.02$ TeV}",
    eprint = "1708.04962",
    archivePrefix = "arXiv",
    primaryClass = "nucl-ex",
    reportNumber = "CMS-HIN-16-001, CERN-EP-2017-186",
    doi = "10.1016/j.physletb.2018.05.074",
    journal = "Phys. Lett. B",
    volume = "782",
    pages = "474--496",
    year = "2018"
}

@article{Kniehl:2007erq,
    author = "Kniehl, Bernd A. and Kramer, Gustav and Schienbein, Ingo and Spiesberger, Hubert",
    title = "{Finite-mass effects on inclusive $B$ meson hadroproduction}",
    eprint = "0705.4392",
    archivePrefix = "arXiv",
    primaryClass = "hep-ph",
    reportNumber = "DESY-07-066, MZ-TH-07-07, LPSC-07-46",
    doi = "10.1103/PhysRevD.77.014011",
    journal = "Phys. Rev. D",
    volume = "77",
    pages = "014011",
    year = "2008"
}

@article{Cacciari:2003uh,
    author = "Cacciari, M. and Frixione, S. and Mangano, M. L. and Nason, P. and Ridolfi, G.",
    title = "{QCD analysis of first $b$ cross-section data at 1.96-TeV}",
    eprint = "hep-ph/0312132",
    archivePrefix = "arXiv",
    reportNumber = "CERN-TH-2003-298, BICOCCA-FT-03-33, GEF-TH-14-2003, UPRF-2003-31",
    doi = "10.1088/1126-6708/2004/07/033",
    journal = "JHEP",
    volume = "07",
    pages = "033",
    year = "2004"
}

@article{Kniehl:2005mk,
    author = "Kniehl, B. A. and Kramer, G. and Schienbein, I. and Spiesberger, H.",
    title = "{Collinear subtractions in hadroproduction of heavy quarks}",
    eprint = "hep-ph/0502194",
    archivePrefix = "arXiv",
    reportNumber = "DESY-05-030, MZ-TH-05-04",
    doi = "10.1140/epjc/s2005-02200-7",
    journal = "Eur. Phys. J. C",
    volume = "41",
    pages = "199--212",
    year = "2005"
}

@article{Cacciari:2003zu,
    author = "Cacciari, Matteo and Nason, Paolo",
    title = "{Charm cross-sections for the Tevatron Run II}",
    eprint = "hep-ph/0306212",
    archivePrefix = "arXiv",
    reportNumber = "UPRF-2003-10, BICOCCA-FT-03-14",
    doi = "10.1088/1126-6708/2003/09/006",
    journal = "JHEP",
    volume = "09",
    pages = "006",
    year = "2003"
}

@article{Qin:2015srf,
    author = "Qin, Guang-You and Wang, Xin-Nian",
    editor = "Wang, Xin-Nian",
    title = "{Jet quenching in high-energy heavy-ion collisions}",
    eprint = "1511.00790",
    archivePrefix = "arXiv",
    primaryClass = "hep-ph",
    doi = "10.1142/S0218301315300143",
    journal = "Int. J. Mod. Phys. E",
    volume = "24",
    number = "11",
    pages = "1530014",
    year = "2015"
}

@article{Qin:2009uh,
    author = "Qin, G. -Y. and Majumder, A. and Song, Huichao and Heinz, U.",
    title = "{Energy and momentum deposited into a QCD medium by a jet shower}",
    eprint = "0903.2255",
    archivePrefix = "arXiv",
    primaryClass = "nucl-th",
    doi = "10.1103/PhysRevLett.103.152303",
    journal = "Phys. Rev. Lett.",
    volume = "103",
    pages = "152303",
    year = "2009"
}

@article{Casalderrey-Solana:2020rsj,
    author = "Casalderrey-Solana, Jorge and Milhano, Jos\'e Guilherme and Pablos, Daniel and Rajagopal, Krishna and Yao, Xiaojun",
    title = "{Jet Wake from Linearized Hydrodynamics}",
    eprint = "2010.01140",
    archivePrefix = "arXiv",
    primaryClass = "hep-ph",
    reportNumber = "MIT-CTP/5246",
    doi = "10.1007/JHEP05(2021)230",
    journal = "JHEP",
    volume = "05",
    pages = "230",
    year = "2021"
}

@article{Djordjevic:2004nq,
    author = "Djordjevic, Magdalena and Gyulassy, Miklos and Wicks, Simon",
    title = "{The Charm and beauty of RHIC and LHC}",
    eprint = "hep-ph/0410372",
    archivePrefix = "arXiv",
    doi = "10.1103/PhysRevLett.94.112301",
    journal = "Phys. Rev. Lett.",
    volume = "94",
    pages = "112301",
    year = "2005"
}

@article{PHENIX:2010cfl,
    author = "Adare, A. and others",
    collaboration = "PHENIX",
    title = "{Azimuthal correlations of electrons from heavy-flavor decay with hadrons in p+p and Au+Au collisions at $\sqrt{s_\mathrm{NN}}=200$ GeV}",
    eprint = "1011.1477",
    archivePrefix = "arXiv",
    primaryClass = "nucl-ex",
    doi = "10.1103/PhysRevC.83.044912",
    journal = "Phys. Rev. C",
    volume = "83",
    pages = "044912",
    year = "2011"
}

@article{STAR:2019qbf,
    author = "Adam, J. and others",
    collaboration = "STAR",
    title = "{Measurement of D$^0$-meson + hadron two-dimensional angular correlations in Au+Au collisions at $\sqrt{s_{\rm NN}}$ = 200 GeV}",
    eprint = "1911.12168",
    archivePrefix = "arXiv",
    primaryClass = "nucl-ex",
    doi = "10.1103/PhysRevC.102.014905",
    journal = "Phys. Rev. C",
    volume = "102",
    number = "1",
    pages = "014905",
    year = "2020"
}

@article{CMS:2022btc,
    author = "Tumasyan, Armen and others",
    collaboration = "CMS",
    title = "{Search for medium effects using jets from bottom quarks in PbPb collisions at $\sqrt{s_\mathrm{NN}}=5.02$ TeV}",
    eprint = "2210.08547",
    archivePrefix = "arXiv",
    primaryClass = "hep-ex",
    reportNumber = "CMS-HIN-20-003, CERN-EP-2022-190",
    doi = "10.1016/j.physletb.2023.137849",
    journal = "Phys. Lett. B",
    volume = "844",
    pages = "137849",
    year = "2023"
}

@article{CMS:2013qak,
    author = "Chatrchyan, Serguei and others",
    collaboration = "CMS",
    title = "{Evidence of b-Jet Quenching in PbPb Collisions at $\sqrt{s_\mathrm{NN}}=2.76$  TeV}",
    eprint = "1312.4198",
    archivePrefix = "arXiv",
    primaryClass = "nucl-ex",
    reportNumber = "CMS-HIN-12-003, CERN-PH-EP-2013-228",
    doi = "10.1103/PhysRevLett.113.132301",
    journal = "Phys. Rev. Lett.",
    volume = "113",
    number = "13",
    pages = "132301",
    year = "2014",
    note = "[Erratum: Phys.Rev.Lett. 115, 029903 (2015)]"
}

@article{Mangano:1991jk,
    author = "Mangano, Michelangelo L. and Nason, Paolo and Ridolfi, Giovanni",
    title = "{Heavy quark correlations in hadron collisions at next-to-leading order}",
    reportNumber = "IFUP-TH-32-91, UPRF-91-308, GEF-TH-10-1991A",
    doi = "10.1016/0550-3213(92)90435-E",
    journal = "Nucl. Phys. B",
    volume = "373",
    pages = "295--345",
    year = "1992"
}

@article{Norrbin:2000zc,
    author = "Norrbin, E. and Sjostrand, T.",
    title = "{Production and hadronization of heavy quarks}",
    eprint = "hep-ph/0005110",
    archivePrefix = "arXiv",
    reportNumber = "LU-TP-00-16",
    doi = "10.1007/s100520000460",
    journal = "Eur. Phys. J. C",
    volume = "17",
    pages = "137--161",
    year = "2000"
}

\newpage
\appendix
\section{Supplemental Figures}
\label{sec:Append}

\begin{figure}[!htbp]
\centering
\includegraphics[width=\linewidth]{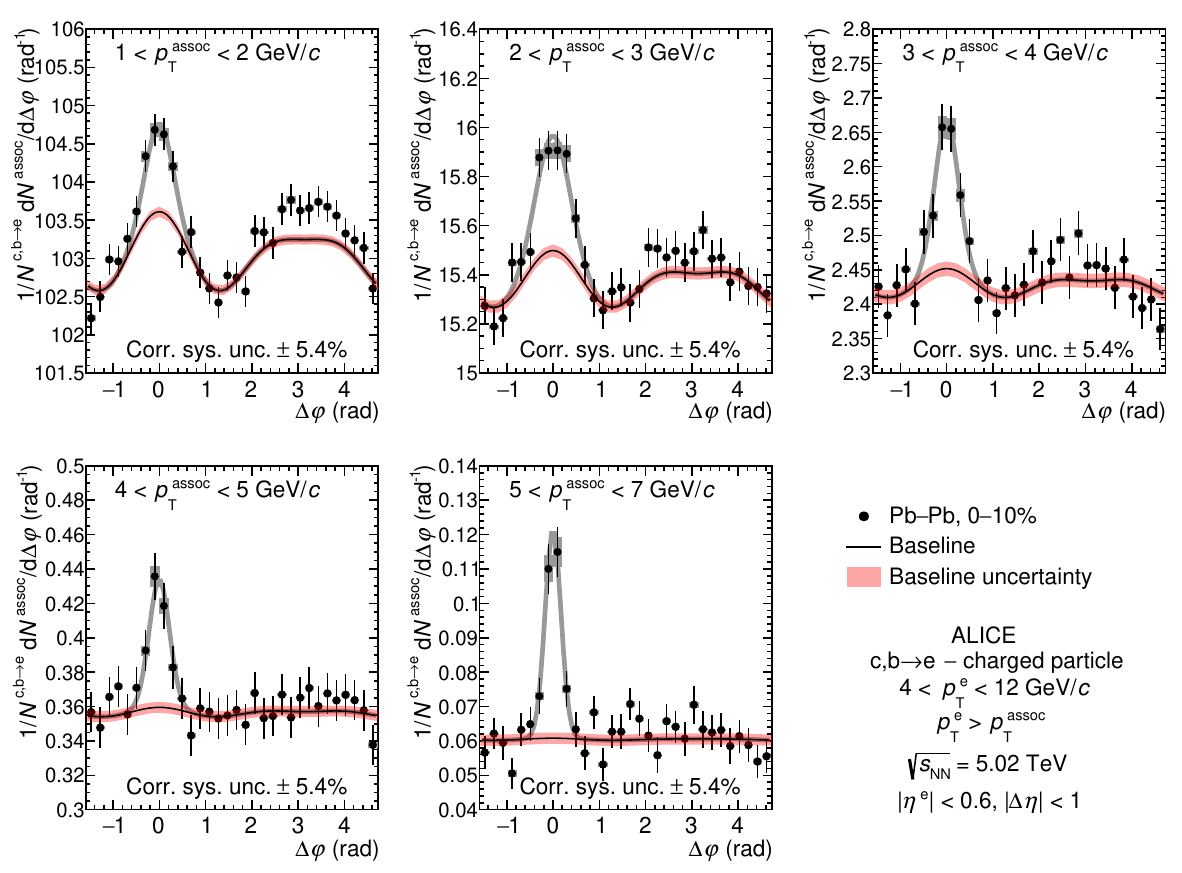}
\caption{Azimuthal-correlation distribution of electrons from heavy-flavor hadron decays and charged particles measured in 0--10\% central Pb--Pb collisions before background subtraction. The distributions are shown in the trigger range of $4 < p_\mathrm{T}^\mathrm{e} < 12$ \gevc and all $p_\mathrm{T}^\mathrm{assoc}$ intervals, where $1 - 2$, $2 - 3$, and $5 - 7$ GeV/$c$ are shown in Fig.~\ref{fig:DeltaPhiPbPb}.}
\label{fig:Delphi_4to12_010_wbackground}
\end{figure}

\begin{figure}[!htbp]
\centering
\includegraphics[width=\linewidth]{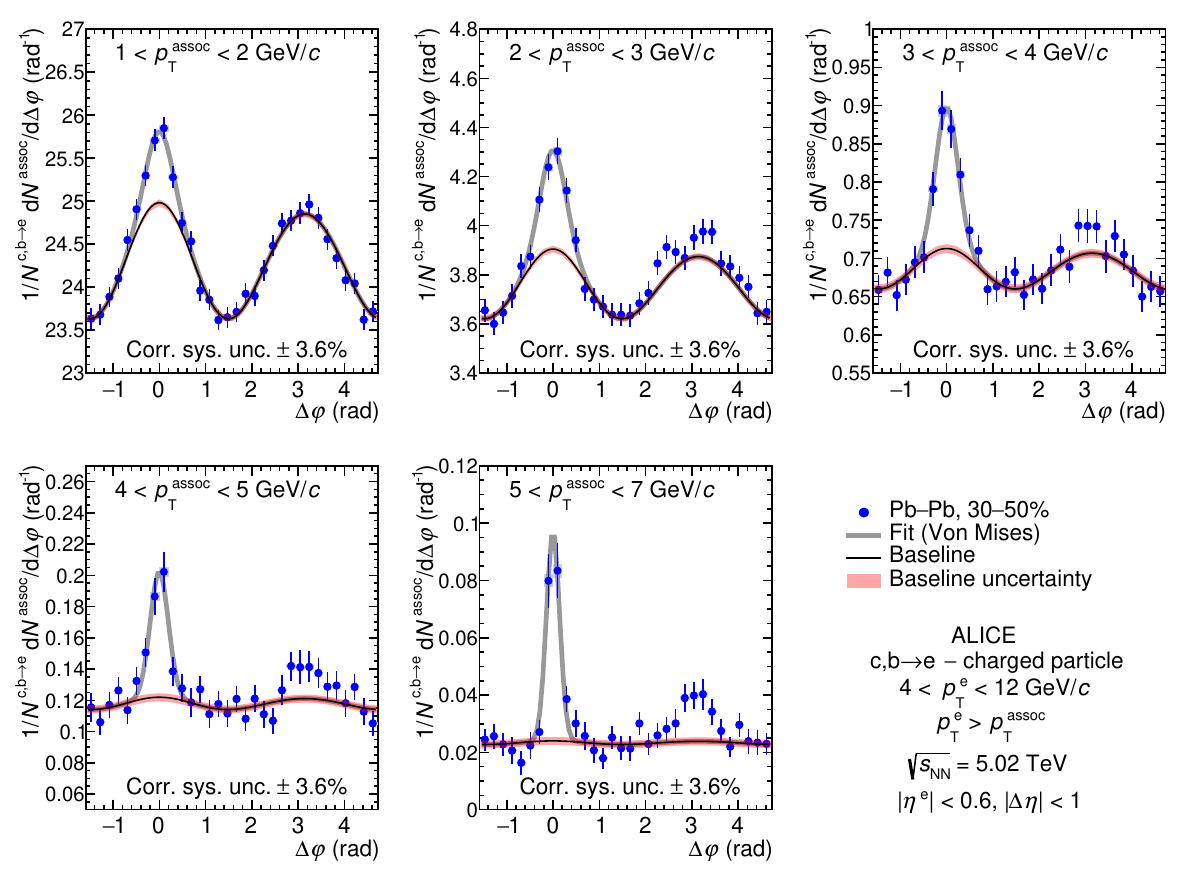}
\caption{Azimuthal-correlation distribution of electrons from heavy-flavor hadron decays and charged particles measured in 30--50\% central Pb--Pb collisions before background subtraction. The distributions are shown in the trigger range of $4 < p_\mathrm{T}^\mathrm{e} < 12$ \gevc and all $p_\mathrm{T}^\mathrm{assoc}$ intervals, where $1 - 2$, $2 - 3$, and $5 - 7$ GeV/$c$ are shown in Fig.~\ref{fig:DeltaPhiPbPb}.}
\label{fig:Delphi_4to12_3050_wbackground}
\end{figure}

\newpage
\section{The ALICE Collaboration}
\label{app:collab}
\begin{flushleft} 
\small

I.J.~Abualrob\,\orcidlink{0009-0005-3519-5631}\,$^{\rm 114}$, 
S.~Acharya\,\orcidlink{0000-0002-9213-5329}\,$^{\rm 50}$, 
G.~Aglieri Rinella\,\orcidlink{0000-0002-9611-3696}\,$^{\rm 32}$, 
L.~Aglietta\,\orcidlink{0009-0003-0763-6802}\,$^{\rm 24}$, 
M.~Agnello\,\orcidlink{0000-0002-0760-5075}\,$^{\rm 29}$, 
N.~Agrawal\,\orcidlink{0000-0003-0348-9836}\,$^{\rm 25}$, 
Z.~Ahammed\,\orcidlink{0000-0001-5241-7412}\,$^{\rm 133}$, 
S.~Ahmad\,\orcidlink{0000-0003-0497-5705}\,$^{\rm 15}$, 
I.~Ahuja\,\orcidlink{0000-0002-4417-1392}\,$^{\rm 36}$, 
Z.~Akbar$^{\rm 81}$
% ...

\section*{Affiliation Notes}

$^{\rm I}$ Deceased\\
$^{\rm II}$ Also at: Max-Planck-Institut fur Physik, Munich, Germany\\
$^{\rm III}$ Also at: Czech Technical University in Prague (CZ)\\
$^{\rm IV}$ Also at: Italian National Agency for New Technologies, Energy and Sustainable Economic Development (ENEA), Bologna, Italy\\
$^{\rm V}$ Also at: Instituto de Fisica da Universidade de Sao Paulo\\
$^{\rm VI}$ Also at: Dipartimento DET del Politecnico di Torino, Turin, Italy\\
$^{\rm VII}$ Also at: Department of Applied Physics, Aligarh Muslim University, Aligarh, India\\
$^{\rm VIII}$ Also at: Institute of Theoretical Physics, University of Wroclaw, Poland\\
$^{\rm IX}$ Also at: Facultad de Ciencias, Universidad Nacional Aut\'{o}noma de M\'{e}xico, Mexico City, Mexico\\

\section*{Collaboration Institutes}

$^{1}$ A.I. Alikhanyan National Science Laboratory (Yerevan Physics Institute) Foundation, Yerevan, Armenia\\
$^{2}$ AGH University of Krakow, Cracow, Poland\\
$^{3}$ Bogolyubov Institute for Theoretical Physics, National Academy of Sciences of Ukraine, Kyiv, Ukraine\\
$^{4}$ Bose Institute, Department of Physics  and Centre for Astroparticle Physics and Space Science (CAPSS), Kolkata, India\\
$^{5}$ California Polytechnic State University, San Luis Obispo, California, United States\\
$^{6}$ Central China Normal University, Wuhan, China\\
$^{7}$ Centro de Aplicaciones Tecnol\'{o}gicas y Desarrollo Nuclear (CEADEN), Havana, Cuba\\
$^{8}$ Centro de Investigaci\'{o}n y de Estudios Avanzados (CINVESTAV), Mexico City and M\'{e}rida, Mexico\\
$^{9}$ Chicago State University, Chicago, Illinois, United States\\
$^{10}$ China Nuclear Data Center, China Institute of Atomic Energy, Beijing, China\\
$^{11}$ China University of Geosciences, Wuhan, China\\
$^{12}$ Chungbuk National University, Cheongju, Republic of Korea\\
$^{13}$ Comenius University Bratislava, Faculty of Mathematics, Physics and Informatics, Bratislava, Slovak Republic\\
$^{14}$ Creighton University, Omaha, Nebraska, United States\\
$^{15}$ Department of Physics, Aligarh Muslim University, Aligarh, India\\
$^{16}$ Department of Physics, Pusan National University, Pusan, Republic of Korea\\
$^{17}$ Department of Physics, Sejong University, Seoul, Republic of Korea\\
$^{18}$ Department of Physics, University of California, Berkeley, California, United States\\
$^{19}$ Department of Physics, University of Oslo, Oslo, Norway\\
$^{20}$ Department of Physics and Technology, University of Bergen, Bergen, Norway\\
$^{21}$ Dipartimento di Fisica, Universit\`{a} di Pavia, Pavia, Italy\\
$^{22}$ Dipartimento di Fisica dell'Universit\`{a} and Sezione INFN, Cagliari, Italy\\
$^{23}$ Dipartimento di Fisica dell'Universit\`{a} and Sezione INFN, Trieste, Italy\\
$^{24}$ Dipartimento di Fisica dell'Universit\`{a} and Sezione INFN, Turin, Italy\\
$^{25}$ Dipartimento di Fisica e Astronomia dell'Universit\`{a} and Sezione INFN, Bologna, Italy\\
$^{26}$ Dipartimento di Fisica e Astronomia dell'Universit\`{a} and Sezione INFN, Catania, Italy\\
$^{27}$ Dipartimento di Fisica e Astronomia dell'Universit\`{a} and Sezione INFN, Padova, Italy\\
$^{28}$ Dipartimento di Fisica `E.R.~Caianiello' dell'Universit\`{a} and Gruppo Collegato INFN, Salerno, Italy\\
$^{29}$ Dipartimento DISAT del Politecnico and Sezione INFN, Turin, Italy\\
$^{30}$ Dipartimento di Scienze MIFT, Universit\`{a} di Messina, Messina, Italy\\
$^{31}$ Dipartimento Interateneo di Fisica `M.~Merlin' and Sezione INFN, Bari, Italy\\
$^{32}$ European Organization for Nuclear Research (CERN), Geneva, Switzerland\\
$^{33}$ Faculty of Electrical Engineering, Mechanical Engineering and Naval Architecture, University of Split, Split, Croatia\\
$^{34}$ Faculty of Nuclear Sciences and Physical Engineering, Czech Technical University in Prague, Prague, Czech Republic\\
$^{35}$ Faculty of Physics, Sofia University, Sofia, Bulgaria\\
$^{36}$ Faculty of Science, P.J.~\v{S}af\'{a}rik University, Ko\v{s}ice, Slovak Republic\\
$^{37}$ Faculty of Technology, Environmental and Social Sciences, Bergen, Norway\\
$^{38}$ Frankfurt Institute for Advanced Studies, Johann Wolfgang Goethe-Universit\"{a}t Frankfurt, Frankfurt, Germany\\
$^{39}$ Fudan University, Shanghai, China\\
$^{40}$ Gangneung-Wonju National University, Gangneung, Republic of Korea\\
$^{41}$ Gauhati University, Department of Physics, Guwahati, India\\
$^{42}$ Helmholtz-Institut f\"{u}r Strahlen- und Kernphysik, Rheinische Friedrich-Wilhelms-Universit\"{a}t Bonn, Bonn, Germany\\
$^{43}$ Helsinki Institute of Physics (HIP), Helsinki, Finland\\
$^{44}$ High Energy Physics Group,  Universidad Aut\'{o}noma de Puebla, Puebla, Mexico\\
$^{45}$ Horia Hulubei National Institute of Physics and Nuclear Engineering, Bucharest, Romania\\
$^{46}$ HUN-REN Wigner Research Centre for Physics, Budapest, Hungary\\
$^{47}$ Indian Institute of Technology Bombay (IIT), Mumbai, India\\
$^{48}$ Indian Institute of Technology Indore, Indore, India\\
$^{49}$ INFN, Laboratori Nazionali di Frascati, Frascati, Italy\\
$^{50}$ INFN, Sezione di Bari, Bari, Italy\\
$^{51}$ INFN, Sezione di Bologna, Bologna, Italy\\
$^{52}$ INFN, Sezione di Cagliari, Cagliari, Italy\\
$^{53}$ INFN, Sezione di Catania, Catania, Italy\\
$^{54}$ INFN, Sezione di Padova, Padova, Italy\\
$^{55}$ INFN, Sezione di Pavia, Pavia, Italy\\
$^{56}$ INFN, Sezione di Torino, Turin, Italy\\
$^{57}$ INFN, Sezione di Trieste, Trieste, Italy\\
$^{58}$ Inha University, Incheon, Republic of Korea\\
$^{59}$ Institute for Gravitational and Subatomic Physics (GRASP), Utrecht University/Nikhef, Utrecht, Netherlands\\
$^{60}$ Institute of Experimental Physics, Slovak Academy of Sciences, Ko\v{s}ice, Slovak Republic\\
$^{61}$ Institute of Physics, Homi Bhabha National Institute, Bhubaneswar, India\\
$^{62}$ Institute of Physics of the Czech Academy of Sciences, Prague, Czech Republic\\
$^{63}$ Institute of Space Science (ISS), Bucharest, Romania\\
$^{64}$ Institut f\"{u}r Kernphysik, Johann Wolfgang Goethe-Universit\"{a}t Frankfurt, Frankfurt, Germany\\
$^{65}$ Instituto de Ciencias Nucleares, Universidad Nacional Aut\'{o}noma de M\'{e}xico, Mexico City, Mexico\\
$^{66}$ Instituto de F\'{i}sica, Universidade Federal do Rio Grande do Sul (UFRGS), Porto Alegre, Brazil\\
$^{67}$ Instituto de F\'{\i}sica, Universidad Nacional Aut\'{o}noma de M\'{e}xico, Mexico City, Mexico\\
$^{68}$ iThemba LABS, National Research Foundation, Somerset West, South Africa\\
$^{69}$ Jeonbuk National University, Jeonju, Republic of Korea\\
$^{70}$ Johann-Wolfgang-Goethe Universit\"{a}t Frankfurt Institut f\"{u}r Informatik, Fachbereich Informatik und Mathematik, Frankfurt, Germany\\
$^{71}$ Korea Institute of Science and Technology Information, Daejeon, Republic of Korea\\
$^{72}$ Laboratoire de Physique Subatomique et de Cosmologie, Universit\'{e} Grenoble-Alpes, CNRS-IN2P3, Grenoble, France\\
$^{73}$ Lawrence Berkeley National Laboratory, Berkeley, California, United States\\
$^{74}$ Lund University Department of Physics, Division of Particle Physics, Lund, Sweden\\
$^{75}$ Nagasaki Institute of Applied Science, Nagasaki, Japan\\
$^{76}$ Nara Women{'}s University (NWU), Nara, Japan\\
$^{77}$ National and Kapodistrian University of Athens, School of Science, Department of Physics , Athens, Greece\\
$^{78}$ National Centre for Nuclear Research, Warsaw, Poland\\
$^{79}$ National Institute of Science Education and Research, Homi Bhabha National Institute, Jatni, India\\
$^{80}$ National Nuclear Research Center, Baku, Azerbaijan\\
$^{81}$ National Research and Innovation Agency - BRIN, Jakarta, Indonesia\\
$^{82}$ Niels Bohr Institute, University of Copenhagen, Copenhagen, Denmark\\
$^{83}$ Nikhef, National institute for subatomic physics, Amsterdam, Netherlands\\
$^{84}$ Nuclear Physics Group, STFC Daresbury Laboratory, Daresbury, United Kingdom\\
$^{85}$ Nuclear Physics Institute of the Czech Academy of Sciences, Husinec-\v{R}e\v{z}, Czech Republic\\
$^{86}$ Oak Ridge National Laboratory, Oak Ridge, Tennessee, United States\\
$^{87}$ Ohio State University, Columbus, Ohio, United States\\
$^{88}$ Physics department, Faculty of science, University of Zagreb, Zagreb, Croatia\\
$^{89}$ Physics Department, Panjab University, Chandigarh, India\\
$^{90}$ Physics Department, University of Jammu, Jammu, India\\
$^{91}$ Physics Program and International Institute for Sustainability with Knotted Chiral Meta Matter (WPI-SKCM$^{2}$), Hiroshima University, Hiroshima, Japan\\
$^{92}$ Physikalisches Institut, Eberhard-Karls-Universit\"{a}t T\"{u}bingen, T\"{u}bingen, Germany\\
$^{93}$ Physikalisches Institut, Ruprecht-Karls-Universit\"{a}t Heidelberg, Heidelberg, Germany\\
$^{94}$ Physik Department, Technische Universit\"{a}t M\"{u}nchen, Munich, Germany\\
$^{95}$ Politecnico di Bari and Sezione INFN, Bari, Italy\\
$^{96}$ Research Division and ExtreMe Matter Institute EMMI, GSI Helmholtzzentrum f\"ur Schwerionenforschung GmbH, Darmstadt, Germany\\
$^{97}$ Saga University, Saga, Japan\\
$^{98}$ Saha Institute of Nuclear Physics, Homi Bhabha National Institute, Kolkata, India\\
$^{99}$ School of Physics and Astronomy, University of Birmingham, Birmingham, United Kingdom\\
$^{100}$ Secci\'{o}n F\'{\i}sica, Departamento de Ciencias, Pontificia Universidad Cat\'{o}lica del Per\'{u}, Lima, Peru\\
$^{101}$ Stefan Meyer Institut f\"{u}r Subatomare Physik (SMI), Vienna, Austria\\
$^{102}$ SUBATECH, IMT Atlantique, Nantes Universit\'{e}, CNRS-IN2P3, Nantes, France\\
$^{103}$ Sungkyunkwan University, Suwon City, Republic of Korea\\
$^{104}$ Suranaree University of Technology, Nakhon Ratchasima, Thailand\\
$^{105}$ Technical University of Ko\v{s}ice, Ko\v{s}ice, Slovak Republic\\
$^{106}$ The Henryk Niewodniczanski Institute of Nuclear Physics, Polish Academy of Sciences, Cracow, Poland\\
$^{107}$ The University of Texas at Austin, Austin, Texas, United States\\
$^{108}$ Universidad Aut\'{o}noma de Sinaloa, Culiac\'{a}n, Mexico\\
$^{109}$ Universidade de S\~{a}o Paulo (USP), S\~{a}o Paulo, Brazil\\
$^{110}$ Universidade Estadual de Campinas (UNICAMP), Campinas, Brazil\\
$^{111}$ Universidade Federal do ABC, Santo Andre, Brazil\\
$^{112}$ Universitatea Nationala de Stiinta si Tehnologie Politehnica Bucuresti, Bucharest, Romania\\
$^{113}$ University of Derby, Derby, United Kingdom\\
$^{114}$ University of Houston, Houston, Texas, United States\\
$^{115}$ University of Jyv\"{a}skyl\"{a}, Jyv\"{a}skyl\"{a}, Finland\\
$^{116}$ University of Kansas, Lawrence, Kansas, United States\\
$^{117}$ University of Liverpool, Liverpool, United Kingdom\\
$^{118}$ University of Science and Technology of China, Hefei, China\\
$^{119}$ University of South-Eastern Norway, Kongsberg, Norway\\
$^{120}$ University of Tennessee, Knoxville, Tennessee, United States\\
$^{121}$ University of the Witwatersrand, Johannesburg, South Africa\\
$^{122}$ University of Tokyo, Tokyo, Japan\\
$^{123}$ University of Tsukuba, Tsukuba, Japan\\
$^{124}$ Universit\"{a}t M\"{u}nster, Institut f\"{u}r Kernphysik, M\"{u}nster, Germany\\
$^{125}$ Universit\'{e} Clermont Auvergne, CNRS/IN2P3, LPC, Clermont-Ferrand, France\\
$^{126}$ Universit\'{e} de Lyon, CNRS/IN2P3, Institut de Physique des 2 Infinis de Lyon, Lyon, France\\
$^{127}$ Universit\'{e} de Strasbourg, CNRS, IPHC UMR 7178, F-67000 Strasbourg, France, Strasbourg, France\\
$^{128}$ Universit\'{e} Paris-Saclay, Centre d'Etudes de Saclay (CEA), IRFU, D\'{e}partment de Physique Nucl\'{e}aire (DPhN), Saclay, France\\
$^{129}$ Universit\'{e}  Paris-Saclay, CNRS/IN2P3, IJCLab, Orsay, France\\
$^{130}$ Universit\`{a} degli Studi di Foggia, Foggia, Italy\\
$^{131}$ Universit\`{a} del Piemonte Orientale, Vercelli, Italy\\
$^{132}$ Universit\`{a} di Brescia, Brescia, Italy\\
$^{133}$ Variable Energy Cyclotron Centre, Homi Bhabha National Institute, Kolkata, India\\
$^{134}$ Warsaw University of Technology, Warsaw, Poland\\
$^{135}$ Wayne State University, Detroit, Michigan, United States\\
$^{136}$ Yale University, New Haven, Connecticut, United States\\
$^{137}$ Yildiz Technical University, Istanbul, Turkey\\
$^{138}$ Yonsei University, Seoul, Republic of Korea\\
$^{139}$ Affiliated with an institute formerly covered by a cooperation agreement with CERN\\
$^{140}$ Affiliated with an international laboratory covered by a cooperation agreement with CERN.\\

\end{flushleft} 
\end{document}